\documentclass[]{aastex631}

\usepackage{amsmath}

\graphicspath{{./}{figures/}}

\begin{document}

\title{Compositional Convection in the Deep Interior of Uranus}

\author{D. J. Hill}
\affiliation{Department of Physics, Drexel University, Philadelphia, PA 19104, USA}

\author{K. M. Soderlund}
\affiliation{Institute for Geophysics, Jackson School of Geosciences, University of Texas at Austin, Austin, TX 78758, USA}

\author{S. L. W. McMillan}
\affiliation{Department of Physics, Drexel University, Philadelphia, PA 19104, USA}

\correspondingauthor{D.J. Hill}
\email{dustin.jay.hill@drexel.edu}

\shorttitle{Deep Compositional Convection in Uranus}

\begin{abstract}
    Uranus and Neptune share properties that are distinct from the other giant planets in the solar system, but they are also distinct from one another, particularly in their relative internal heat flux.
    Not only does Neptune emit about ten times the amount of heat that emitted by Uranus, the relative amount of emitted heat to the energy they absorb from the sun also differs greatly, being comparable at Uranus and the largest of all giant planets at Neptune.
    As a result, it is questionable whether thermal convection occurs within the interior of Uranus.
    However, the presence of an intrinsic magnetic field implies that interior fluid motions must exist.
    Here, we consider compositional convection driven by the release of hydrogen associated with the formation of large organic networks or diamond precipitation in the deep interior.
    We test this hypotheses using a set of numerical rotating convection models where the convective driving is varied between thermal and compositional sources and is sufficiently vigorous to not be strongly constrained by rotation.
    In most cases, we find ice-giant-like zonal flows develop, with three bands characterized by a retrograde equatorial jet and prograde jets at higher latitudes.
    Large-scale circulation cells also develop and lead to heat and mass fluxes that tend to exhibit local maxima along the equatorial plane.
    This similarity between convective flows driven by thermal and compositional buoyancy therefore predict Uranus and Neptune to have similar interior dynamics despite Uranus' minimal internal heat flow and
    may thus explain why both ice giants have comparable magnetic fields.
\end{abstract}

\section{Introduction}

Uranus and Neptune share a number of similarities that have lead them to be considered twin planets, often referred to collectively as the ice giants, differentiating them from the gas giant planets, Jupiter and Saturn.
Uranus and Neptune have similar masses (14.5 $M_\oplus$ and 17.1 $M_\oplus$), radii (4.0 $R_\oplus$ and 3.9 $R_\oplus$), rotational periods ($1.012 \times 10^{-4}$ s$^{-1}$ and $1.083 \times 10^{-4}$ s$^{-1}$), atmospheric compositions (dominated by hydrogen/helium and enriched in heavier elements), three-jet zonal winds, non-dipole dominated magnetic fields, and measured gravitational moments \citep{jacobson2009orbits,jacobson2014orbits,williams2015nasa}.
However, despite these similarities, there are important differences between the two planets that may give insights into the evolution of planetary systems.
For example, Uranus has a lower mean density than Neptune (1200 kg m$^{-3}$ compared to 1600 kg m$^{-3}$) and a uniquely extreme axial tilt of 97.8$^\circ$, compared to the more Earth-like tilt of 28.3$^\circ$ of Neptune.
Though both Uranus and Neptune have colder effective temperatures than the other giant planets, Neptune emits the greatest amount of heat relative to that received through insolation while Uranus has the lowest among the giant planets \citep{pearl1990albedo,pearl1991albedo}.
As will be described below, we hypothesize that compositional convection---rather than thermal---in Uranus' deep interior may explain the dynamical similarities of the ice giant planets.

In this paper, we first discuss the background and scientific rationale for studying different drivers of deep convection in ice giant interiors.
Section~\ref{sec:regimes} further discusses the dynamical regimes of convective motions that may be expected in giant planet interiors.
Section~\ref{sec:methods} then details the numerical model of rotating convection in a spherical shell used to test thermal and/or compositional buoyancy as the origin of flows in the deep interior of Uranus.
The velocity and thermodynamic fields found in our numerical experiments are described in Section~\ref{sec:results}, and the underlying processes are discussed in Section~\ref{sec:analysis}.
Section~\ref{sec:discussion} extrapolates these results to ice giant interiors.

\subsection{Internal structure, composition, and heat flow}

The distribution of mass in planetary interiors is measured by gravitational moments, which are non-unique with respect to both the radial dependence of density as well as composition and temperature.
Traditionally, internal structures of Uranus and Neptune are modeled with three distinct layers: a solid rocky core, a mantle consisting primarily of molecular ices (e.g., water, methane, and ammonia), and an atmosphere consisting of molecular hydrogen and helium and other gases such as methane and ammonia \citep[e.g.,][]{hubbard1991interior,podolak1991models,nettelmann2013new}.
The ratio of rock to ice remains an open question, and the measured gravitational moments, $J_2$ and $J_4$, may be explained by either an ice-dominated or a rock-dominated composition \citep[e.g.,][]{teanby2020neptune}. There may also be a continuous variation of density and concentration of heavy elements through the interior rather than distinct phase transitions separating layers \citep{fortney2010giant,helled2010interior,movshovitz2020promise,vazan2020explaining}.

Nonetheless, the interiors of Uranus and Neptune are thought to contain a mixture of materials containing carbon, nitrogen, and oxygen \citep{atreya2020deep}.
Hydrocarbons, primarily methane, are present in the atmospheres of the ice giants, forming both clouds and hazes in the upper atmosphere, which obscure deeper regions of the atmosphere, making deep mixing ratios of carbon to hydrogen difficult to measure \citep{lindal1987atmosphere,baines1995abundances}.
Measurements of the abundances of atmospheric methane show that both planets are substantially enriched in carbon, approaching 100 times solar ratios of carbon-to-hydrogen \citep{atreya2020deep}.
At high pressures and temperatures, methane has been shown to dissociate into diamond (carbon) and hydrogen, rather than forming a superionic solid \citep{ross1981ice,kraus2017formation}.
Simulations indicate that mixtures of methane, ammonia, and water also enter such a dissociative state \citep{chau2011chemical,naumova2020unusual}.

While atmospheric abundances of methane and other carbon-rich molecules can be measured through ground based observations, the atmospheric abundances of ammonia and water poorly known \citep{mousis2020key}.
Water has been studied under planetary interior conditions both experimentally and theoretically, predicting that both electrically conductive ionic fluids and superionic solids may occur within the ice giants \citep{french2016ab,millot2018experimental}.
Ammonia and mixtures of ammonia and water exhibit similar states of matter \citep{cavazzoni1999superionic, ravasio2021metallization,bethkenhagen2017planetary}.

\begin{figure}
	\fig{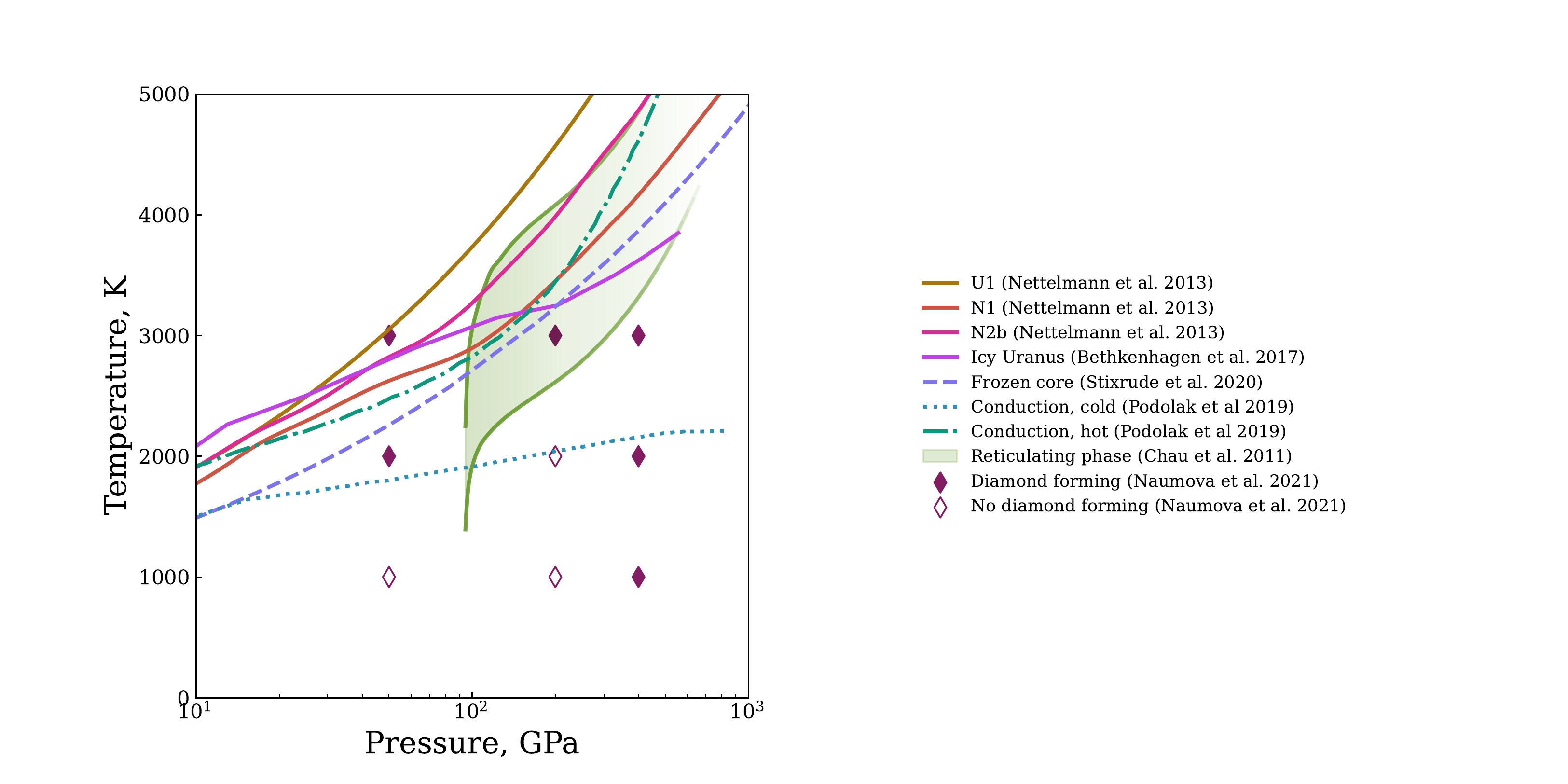}{\linewidth}{}
	\vspace*{-10mm}
	\caption{Pressure-temperature diagrams for both materials thought to be present in the ice giant interiors as well as specific internal structure models of Uranus and Neptune that feature both fully convective models \citep{nettelmann2013new,bethkenhagen2017planetary} and models that include significant non-adiabatic regions \citep{podolak2019effect,stixrude2020thermal}. The green shaded area represents the chemically active reticulating phase of synthetic Uranus, though the boundaries of this region are approximate \citep{chau2011chemical}.  The region fades with increasing temperature, as the extent in temperature of this region of phase space are also unknown.  Diamonds represent the states of a mixture consisting of water, methane, and ammonia in a ratio of 56.5:32.5:11, where filled symbols denote where demixing into diamonds occurs and empty symbols denote where this demixing does not occur \citep{naumova2020unusual}.}
	\label{fig:phase_diagram}
\end{figure}

Figure~\ref{fig:phase_diagram} shows the pressure-temperature diagram for several models of the interiors of Uranus and Neptune and relevant chemical states that may occur in their interiors.
``Synthetic Uranus" approximates the interior composition by a ratio of hydrogen to carbon to nitrogen to oxygen (H:C:N:O) of 28:4:1:7, which has been shown to exist in different phases of matter and to become electrically conductive with increasing pressure and temperature \citep{nellis1988nature,chau2011chemical,guarguaglini2019laser}.
\textit{Ab initio} simulations predict that synthetic Uranus enters what is referred to as a reticulating phase, where C--C and C--N bonds become energetically favorable and form networks within the material \citep{chau2011chemical}.
In a similar mixture of water, ammonia, and methane, carbon may demix from the solution to form diamonds \citep{naumova2020unusual}.
In both cases, hydrogen may be released as a product of the reactions.
This release of light components may then be a potential source of compositional buoyancy, similar to the release of light elements as Earth's inner core solidifies.

Besides composition, the thermal structures of the ice giants are similarly not well constrained, which are reflected by the inclusion of both adiabatic and non-adiabatic interior models in Figure~\ref{fig:phase_diagram} \citep{nettelmann2013new,podolak1991models,stixrude2020thermal}.
Magnetic fields were discovered at Uranus and Neptune by the \textit{Voyager 2} mission \citep{ness1986magnetic,ness1989magnetic}.
The existence of planetary magnetic fields can be explained by dynamo theory, which states that a planetary magnetic field is sustained against diffusion by induction caused by complex motion within an electrically conductive fluid in the planet’s deeper interior \citep{luhr2018magnetic}.
For this reason, we can infer the existence of convective motion in, at least a part of, the interiors of the ice giants.
However, adiabatic models of Uranus and Neptune that assume fully convective interiors throughout their history are unable to correctly predict the thermal properties of either planet in the present, suggesting that the planetary interiors are only partially convective or that their thermal behaviors are time-dependent \citep{scheibe2019thermal}.

Thermal properties of the giant planets are summarized in Table \ref{tab:thermal}.
The majority of energy emitted by Neptune originates from the interior, compared to only about 6\% for Uranus \citep{pearl1990albedo,pearl1991albedo}.
The uncertainty in \textit{Voyager 2} measurements is quite large, and it is therefore possible that no internal heat contribution was measured at Uranus.
The thermal properties of planets have traditionally been assumed to be constant in time over decadal timescales; however, measurements made by \textit{Cassini} of Saturn revealed a measurable decrease in emitted power from the period of 2005 to 2009, which have raised questions as to the temporal variation of thermal properties of the other giant planets  \citep{li2010saturn}.
Furthermore, the instruments aboard the \textit{Cassini} spacecraft were able to make measurements of Jupiter that revealed that more heat is emitted than had been previously measured by the \textit{Pioneer} or \textit{Voyager} missions \citep{li2018less}.
In contrast, the temporal variability of Uranus' and Neptune's thermal properties are unknown because subsequent measurements have not yet been made.

The low luminosity of Uranus may be accounted for in several possible ways, such as a continuous compositional gradient within the interior or the existence of a thermal boundary layer that could prevent convective transport of heat to the surface \citep{podolak1991models,nettelmann2016uranus,vazan2019explaining}.
Condensation of heavier molecules, such as methane or water, out of hydrogen dominated atmospheres has also been suggested as a mechanism that may inhibit thermal convection \citep{guillot1995condensation,cavalie2017thermochemistry,friedson2017inhibition,leconte2017condensation,markham2021constraining}. 
If convection is inhibited in both Uranus and Neptune, but condensation only occurs in the atmosphere of Neptune, then latent heat associated with this process may warm the atmosphere and prevent it from cooling to a temperature similar to Uranus \citep{kurosaki2017acceleration}.
Additionally, the difference in heating may be caused by demixing of water and hydrogen in the deep interior that occurs in Neptune, but not in Uranus \citep{bailey2021thermodynamically}.
Regions of the deep interior may experience heat transport primarily by conduction, which also predicts a smaller amount of heat reaching the surface \citep{podolak2019effect}.
The low luminosity of Uranus is also consistent with a frozen-core, which is able to sequester primordial heat as the freezing process continues by inhibiting convection throughout the deep interior of the planet \citep{stixrude2020thermal}.
Here, we hypothesize that Uranus' deep interior is unstable convection that is driven either partly or entirely by compositional buoyancy because of the light element release at depth and likely non-adiabaticity of the planet's interior.
This could differentiate Uranus from Neptune, whose significant internal heat flux suggests thermally driven convection.

\begin{deluxetable}{lCCCCCCCCC}
    \tablecaption{Thermal properties and estimates of non-dimensional parameters of the giant planets.  Values for Jupiter are reported in \cite{li2018less}, for Uranus in \cite{pearl1990albedo}, and for Neptune in \cite{pearl1991albedo}. Approximate values of viscosity and thermal diffusivity are based on first principles calculations in \cite{french2012ab} and \cite{french2019viscosity}.  Uranus and Neptune are assumed to consist of water, therefore other properties are estimated using the equation of state presented in \cite{mazevet2019ab}.  Thermal Rayleigh numbers are estimated following \cite{soderlund2019ocean}.  \label{tab:thermal}}
    \tablehead{
        \colhead{} & \colhead{Energy} & \colhead{Internal} & \colhead{Internal} & \colhead{} & \colhead{} & \colhead{} & \colhead{} & \colhead{} & \colhead{} \\
        \colhead{Planet} & \colhead{balance} & \colhead{heat source} & \colhead{heat flux} & \colhead{$\chi$} & \colhead{$Ek$} & \colhead{$Pr$} & \colhead{$Ra$} & \colhead{$Ro_C$} \\
        \colhead{} & \colhead{} & \colhead{($\times 10^{15} $W)} & \colhead{ (W m$^{-2}$)} &
    }
    \startdata
        Jupiter & 2.23 \pm 0.03 & 460 \pm 10 & 7.49 \pm 0.16 & 0.3 - 0.5 & 10^{-19}-10^{-18} & 0.01-0.1 & 10^{28}-10^{31} & 10^{-4}-10^{-3} \\
        Uranus  & 1.06 \pm 0.08 & 0.25 \pm 0.36 & 0.029 \pm 0.042 & 0.3-0.7 & 10^{-18}-10^{-17} & 0.1 - 1 & \lesssim 10^{31} & 10^{-3}-10^{-2} \\
        Neptune & 2.61 \pm 0.28 & 2.35 \pm 0.44 & 0.308 \pm 0.058 & 0.3-0.7 & 10^{-18}-10^{-17} & 0.1 - 1 & 10^{27} - 10^{31} & 10^{-3}-10^{-2} \\
    \enddata
\end{deluxetable}

\subsection{Velocity fields}

\begin{figure}
    \fig{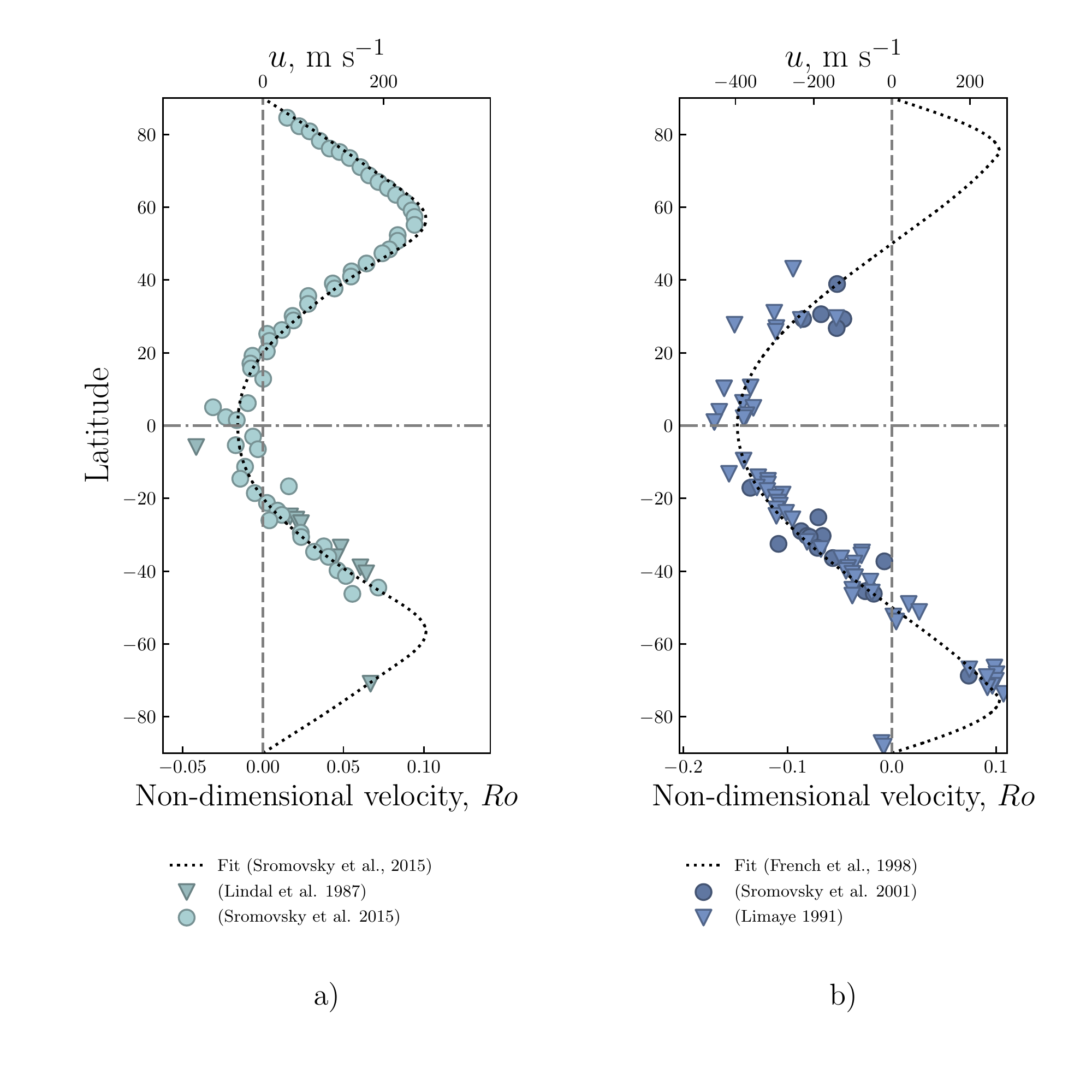}{.75\linewidth}{}
    \vspace*{-10mm}
	\caption{Zonal velocity as a function of latitude at the 1 bar level for a) Uranus and b) Neptune.  Winds are expressed both in terms of physical units (m s$^{-1}$) and non-dimensionally in terms of the  Rossby number, $Ro = u/r_o \Omega$.  Measurements were made by \textit{Voyager 2} and Earth-based observations \citep{lindal1987atmosphere,limaye1991winds,sromovsky2015high,sromovsky2001neptune}.  Empirical fits are provided by \cite{sromovsky2015high} and \cite{french1998neptune} for Uranus and Neptune, respectively.  Grey lines mark the equator (dash-dot) and the zero wind velocity (dashed).
	}
	\label{fig:zonal_winds}
\end{figure}

As shown in Figure~\ref{fig:zonal_winds}, Uranus and Neptune have surface winds that are characterized by three zonal (east-west) jets, with retrograde (westward) winds at the equator and prograde (eastward) winds at higher latitudes \citep{mousis2018scientific}.
In addition to physical units, velocities are expressed in terms of the  he Rossby number, $Ro$, the ratio of inertial to Coriolis forces:
\begin{equation}
    Ro = \frac{u}{r_o \Omega} \label{eq:rossby},
\end{equation}
\noindent where $u$ is velocity, $\Omega$ is rate of rotation of system, and $r_o$ is outer radius.
The Rossby number is dependent on length scale, and we choose $r_o$ so that these $Ro$ values serve as a lower bound; this length scale will be revisited in later sections.
Equatorial winds on Neptune are more intense, reaching 400 m s$^{-1}$ compared to $\sim$ 100 m s$^{-1}$ on Uranus; the extent of these winds is also wider on Neptune, extending between $\pm 20^{\circ}$ on Uranus compared to $\pm 50^{\circ}$ on Neptune.
However, these measurements are calculated on the basis of the rotational periods measured by \textit{Voyager 2}, and modified rotational periods can result in similar winds of $\sim 200$ m s$^{-1}$ at both Uranus and Neptune \citep{helled2010uranus}.
The prograde winds tend to be similar in intensity, $\sim$250 m s$^{-1}$ on Uranus and $\sim$300 m s$^{-1}$ on Neptune, but these prograde bands tend to be broader on Uranus than on Neptune.
It is unknown how deep these winds extend into the interior, although gravity measurements and Ohmic dissipation estimates suggest they are confined to about the top 1-2 thousand km in both planets \citep{kaspi2013atmospheric,soyuer2020constraining}. 

Zonal winds may develop from a number of possible physical processes that typically rely on being either driven by insolation or by internal heat.
Jupiter-like or ice giant-like winds can develop in a thin weather layer through a turbulent inverse energy cascade \citep{cho1996morphogenesis} or moist convection  \citep{lian2009generation}.
Deep-convection can similarly generate zonal flows that have a prograde equatorial jet with alternating jets at higher latitudes (i.e. jupiter-like) in systems strongly constrained by rotation or three jets with retrograde equatorial flow (i.e. ice giant-like) when the system is not rotationally dominated \citep[e.g.,][]{busse1976simple,aurnou2007effects,soderlund2013turbulent,gastine2013solar}. While these latter studies have focused on thermal buoyancy, we hypothesize that compositional convection in the deep interior may also be a viable mechanism to generate Uranus' zonal flows.

\section{Regimes of rotating convection \label{sec:regimes}}

The properties of convective motion are dependent on the relative influence of rotation and the physical characteristics of the fluid, and varying these parameters can result in different regimes of rotating convection \citep{gastine2016scaling,cheng2018heuristic,cheng2020laboratory}.
The relative importance of rotation in the system can be quantified using the Ekman number, while the properties of thermal transport can be quantified using the Prandtl and Rayleigh numbers:
\begin{align}
	Ek & = \frac{\nu}{\Omega D^2} \sim \frac{\text{Viscous diffusion}}{\text{Coriolis force}}, \label{eq:Ekman} \\
	Pr & = \frac{\nu}{\kappa} \sim \frac{\text{Momentum diffusivity}}{\text{Thermal diffusivity}}, \label{eq:Prandtl} \\
	Ra & =  \frac{\alpha g \Delta T D^3}{\nu \kappa} \sim \frac{\text{Thermal buoyancy}}{\text{Diffusion}} \label{eq:Rayleigh}.
\end{align}
\noindent Here, $D$ is length scale, $\nu$ is kinematic viscosity, $\kappa$ is thermal diffusivity, $\alpha$ is the coefficient of thermal expansion, $g$ is gravitational acceleration, and $\Delta T$ is the superadiabatic temperature contrast that drives convection.
Convective heat transfer efficiency of the system is characterized by the ratio of total to conductive heat flux, known as the Nusselt number:
\begin{equation}
	Nu = \frac{q D}{\rho c_p \kappa \Delta T} \label{eq:nusselt},
\end{equation}
\noindent where $q$ is heat flux, $\rho$ is density, and $c_p$ is specific heat capacity.
Trends in $Nu$ behavior as a function of $Ra$, $Pr$, and $Ek$ are often used to identify different convective regimes. 

\cite{cheng2018heuristic} describe six regimes, which are defined by the strength of convective forcing: sub-critical, columnar, plumes, geostrophic turbulence, unbalanced boundary layer, and non-rotating heat transfer.
In rotating systems, the onset of convection occurs once the Rayleigh number reaches a critical value, $Ra_C \sim Ek^{-4/3}$.
As the convective forcing is increased, the dynamics of the system and properties of the heat transfer will be determined by the influence of rotation.
When buoyant forcing is weak, rotation is able to impose structure in the fluid that is constant on surfaces parallel to the axis of rotation.
As the convective forcing is increased, the organization imposed by rotation begins to break down in the so-called unbalanced boundary layer regime and then ultimately the system behaves as if it is non-rotating.

The thermal convective regime for a planetary system can then be predicted using estimates for $Ek$, $Pr$, and $Ra$. While $Ek$ and $Pr$ are relatively straight forward, the Rayleigh number and $\Delta T$ in particular is more challenging. Following \cite{soderlund2019ocean}, the superadiabatic temperature contrast can be obtained by using $Nu-Ra$ scaling laws for rapidly rotating ($\Delta T_{RR}$) and non-rotating heat transfer ($\Delta T_{NR}$) as upper and lower bounds for a given heat flux $q$:

\begin{equation}
    \Delta T_{NR} = 7.3 \left( \frac{q^3 \nu}{\alpha g \rho^3 c_p^3 \kappa^2} \right)^{1/4} \label{eq:temp_nr},
\end{equation}

\begin{equation}
    \Delta T_{RR} = 2.1 \left( \frac{\Omega^4 q^2 \kappa D}{\rho^2 c_p^2 \nu \alpha^3 g^3} \right)^{1/5} \label{eq:temp_rr}.
\end{equation}

\noindent Table \ref{tab:thermal} contains estimates of the non-dimensional parameters that describe the physical properties of the giant planets, assuming a water-dominated interior composition.
At high pressures and temperatures relevant to the ice giant interiors, water is predicted to have a kinematic viscosity of $\nu \sim 3 \times 10^{-8}$ to $10^{-7}$ m$^2$ s$^{-1}$ and Prandtl numbers between 0.1 and 1 \citep{french2019viscosity}.
If we allow the length scale to be between 30\% and 70\% the planetary radius \citep{helled2010interior,nettelmann2013new}, the Ekman numbers are between $10^{-18}$ and $10^{-17}$.
The thermal expansion coefficient can be estimated as $\alpha \sim 10^{-4}$ K$^{-1}$ \citep{mazevet2019ab}, and the value of $\rho c_p$ is estimated to be between $8.7 \times 10^6$ and $2.6 \times 10^7$ J m$^3$ K$^{-1}$ \citep{french2019viscosity}.
The rates of rotation and gravitational field are assumed to match planetary values of $\Omega \sim 10^{-4}$ s$^{-1}$ and $g \sim 10$ m s$^{-2}$. 
Estimates for the properties of the interior of Jupiter are estimated based on the equations of state of hydrogen-helium mixtures, given a viscosity of $\nu \sim \times 10^{-7}$ to $10^{-6}$ m$^2$ s$^{-1}$ and a length scale between 30\% and 50\% the radius of Jupiter results in an Ekman number between $10^{-19}$ and $10^{-18}$ \citep{french2012ab,wahl2017comparing}.
To estimate the Rayleigh number, we let $\rho c_p \approx 3 \times 10^8$ J m$^3$ K$^{-1}$ and allow the Prandtl number to be between 0.01 and 0.1 \citep{chabrier2019new}.
Using these properties, the range of internal heat fluxes from Table~\ref{tab:thermal}, and the temperature contrasts from equations \ref{eq:temp_nr} and \ref{eq:temp_rr}, we estimate Rayleigh numbers of $Ra \sim 10^{27}$ to $10^{31}$ for Neptune based on the mean value of heat in Table~\ref{tab:thermal} and $Ra \lesssim 10^{31}$ for Uranus taking the possibility of zero internal heat source into account.

\begin{figure}
    \fig{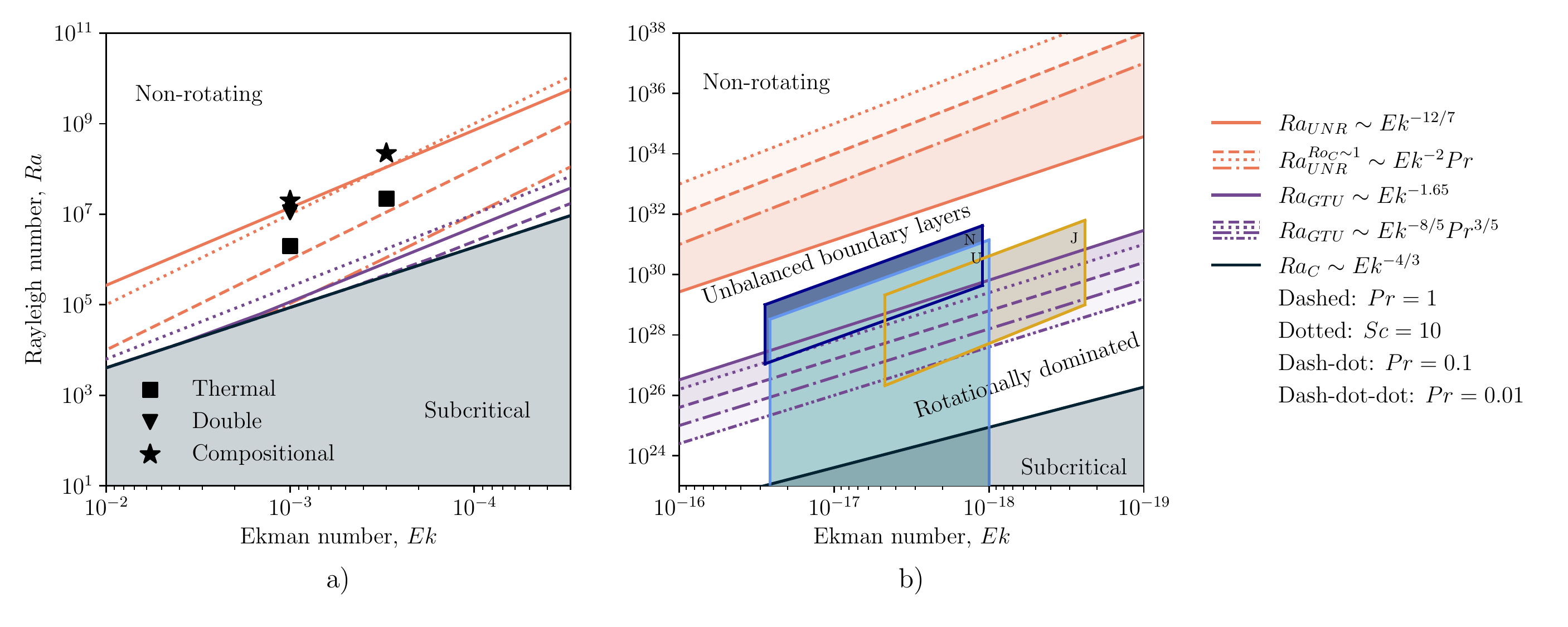}{\linewidth}{}
    \caption{Convective regime diagram for a) the models in this work and b) estimates for the giant planets found in Table~\ref{tab:thermal}, represented by the shaded boxes for Uranus (U), Neptune (N), and Jupiter (J).  In a), the Rayleigh number of the double diffusive case is expressed as the total Rayleigh number.  The grey region indicates Rayleigh numbers that cannot sustain convection at a given Ekman number ($Ra_C$).  The purple lines and shaded bands represent transitions between rotationally dominated regimes of convection and the unbalanced boundary layer regime ($Ra_{GTU}$), and red lines and shaded bands represent transition between unbalanced boundary layers and non-rotating convection ($Ra_{UNR}$).}
    \label{fig:convective_regimes}
\end{figure}

Estimates of the Rayleigh and Ekman numbers for the giant planets are shown in the context of these regimes of rotating convection in Figure~\ref{fig:convective_regimes}b.
Uranus and Neptune fit within the weakly-rotating unbalanced boundary layer regime and the transitional regime between unbalanced boundary layers and rotationally dominated convection.
In contrast, Jupiter appears to be confined mostly to the transitional regime between rotationally dominated and weakly rotating regimes.
When the Rayleigh numbers are calculated on the basis of the mean heat flux, Jupiter is confined to this transitional region, and only enters into the unbalanced boundary layer regime as larger values of the heat flux within one standard error of the mean are considered.
However, these only take thermal convection into account, and any compositional contributions to convection in these planets is unconstrained by observations, meaning that either Neptune or Uranus may have a higher total Rayleigh number, which would lift them further towards the non-rotating regime.
This lack of a strong rotational constraint will guide the choice of input parameters for the models presented in this work.
Choices for our models are shown in Figure~\ref{fig:convective_regimes}a.
At high Ekman number, these regimes are less well differentiated, so we chose parameters in the non-rotating regime to reflect this weak rotational constraint.

\section{Methods \label{sec:methods}}

\subsection{Model}

In this study, we investigate the effect of varying the type of convective driving on the resulting dynamics of the system.
Given the significantly different thermal behaviors displayed between Uranus and Neptune, we will consider cases that are purely thermal, purely compositional, and driven by both thermal and compositional buoyancy (i.e. double-diffusive convection, albeit with both unstable gradients). 

\begin{figure}
    \fig{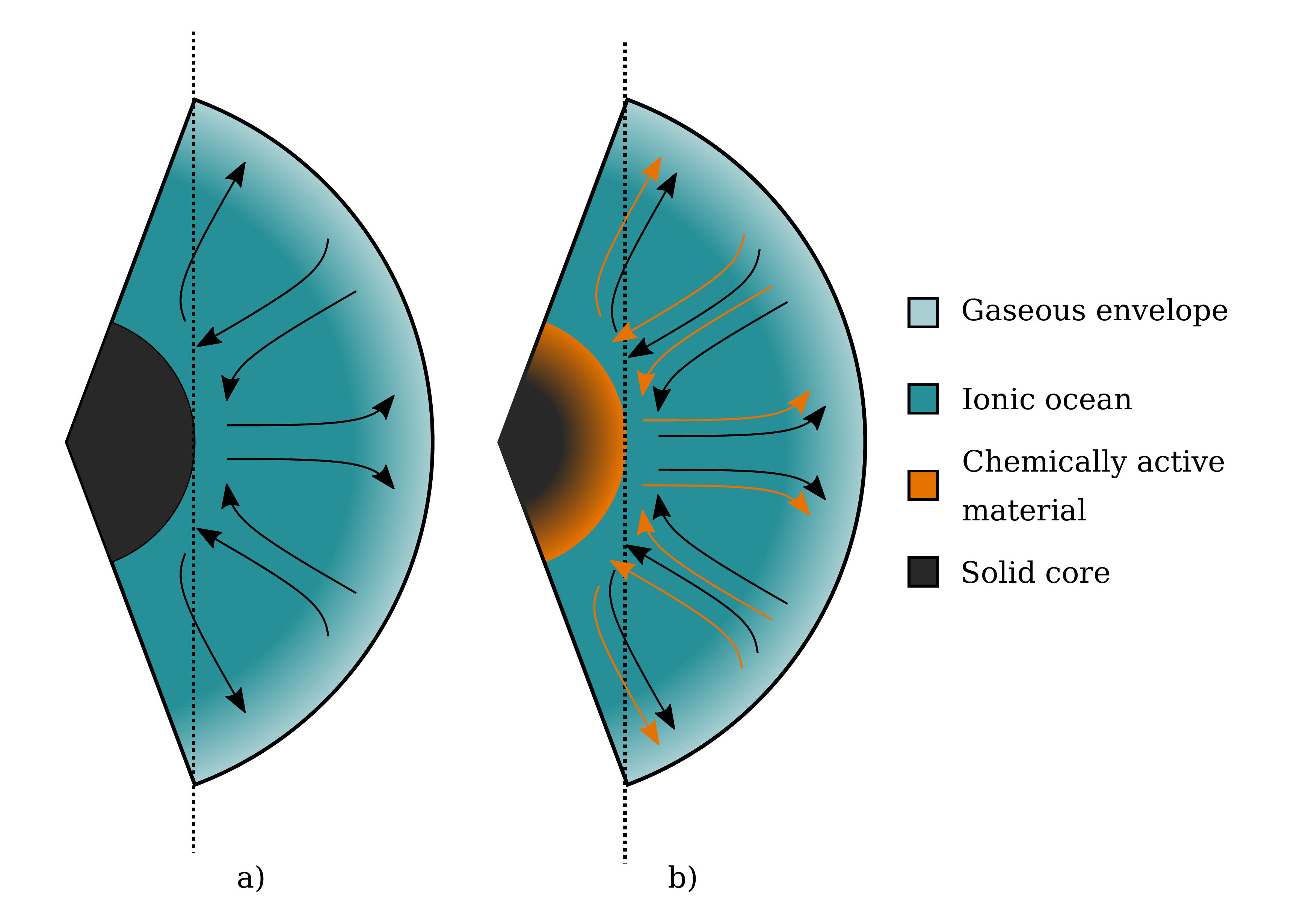}{.6\linewidth}{} 
    \caption{Internal structure schematics for a) a traditional three-layer model where convection is driven thermally and b) a layered structure with a chemically active region of the core where there is an additional compositional source of buoyancy. The dotted vertical line indicates the tangent cylinder, which encloses the central core such that it touches the inner boundary only at the equator and is parallel to the axis of rotation.}
    \label{fig:internal_schematic}
\end{figure}

Figure~\ref{fig:internal_schematic} shows schematics of the internal structures assumed in our models.
Convection is assumed to occur in the ionic ocean and gaseous envelope, while the solid core that lies underneath is unable to move on a timescale relevant to dynamics of the system.
While the existence of magnetic fields in the ice giants implies convective motion, using scaling law arguments, the Lorentz force is estimated to be several orders of magnitude weaker than the Coriolis force in ice giant interiors \citep{soderlund2015competition}.
As a result, the magnetic fields are not expected to modify the convective flows significantly, which justifies our use of hydrodynamic models.

We model thermo-compositional convection of a Boussinesq fluid confined within a spherical shell that rotates at a constant rate about the ${\bf \hat{z}}$-axis. The fluid's behavior is governed by a system of dimensionless equations that describe the time evolution of velocity $\textbf{u}$, temperature $T$, and concentration of light components dissolved in the fluid $\xi$ under the assumption of incompressibility:

\begin{align}
	\frac{\partial \textbf{u}}{\partial t} & = - \textbf{u} \cdot \nabla \textbf{u} -\nabla p - \frac{2}{Ek} \hat{\textbf{z}} \times \textbf{u} + \frac{Ra}{Pr} T \textbf{g} + \frac{Ra_\xi}{Sc} \xi \textbf{g} + \nabla^2 \textbf{u}  \label{eq:navier_stokes}, \\
	\frac{\partial T}{\partial t} & = - \textbf{u} \cdot \nabla T + \frac{1}{Pr} \nabla^2 T  \label{eq:heat}, \\ 
	\frac{\partial \xi}{\partial t} & = - \textbf{u} \cdot \nabla \xi + \frac{1}{Sc} \nabla^2 \xi  \label{eq:mass},
\end{align}
\begin{equation}
    \nabla \cdot \textbf{u} = 0  \label{eq:boussinesq}.
\end{equation}
\noindent The gravitational field $\textbf{g}$ is assumed to be proportional to the radius: $\textbf{g} = g_o \textbf{r}/r_o$, where $g_o$ is gravitational acceleration at the outer boundary. 
The equations are non-dimensionalized using shell thickness $D$ as the characteristic length scale of the system and viscous diffusion time $\tau_{\nu}=D^2/\nu$ as the characteristic time scale, while temperature and the concentration are expressed in terms of the contrast between the inner and outer boundaries, $\Delta T$ and $\Delta \xi$, respectively. 

In addition to the Ekman, Prandtl and Rayleigh number, the system is defined by three other non-dimensional input parameters: Schmidt number $Sc$, compositional Rayleigh number $Ra_\xi$, and shell geometry $\chi$:

\begin{align}
	Sc & = \frac{\nu}{\kappa_\xi} \sim \frac{\text{Momentum diffusivity}}{\text{Mass diffusivity}} \label{eq:Schmidt},  \\
	Ra_\xi & = \frac{\alpha_\xi g_o \Delta \xi D^3}{\nu \kappa_\xi} \sim \frac{\text{Compositional buoyancy }}{\text{Diffusion}} \label{eq:ChemicalRayleigh},  \\
	\chi & = \frac{r_i}{r_o} \sim \frac{\text{Inner boundary radius}}{\text{Outer boundary radius}} \label{eq:geometry}. 
\end{align}

\noindent Here, $\kappa_\xi$ is diffusivity of light components, and $r_i$ and $r_o$ are the inner and outer shell radii, respectively.

We assume a Boussinesq fluid, where fluid properties are assumed constant except for density anomalies associated with temperature and composition perturbations. 
Density is equal to a mean value plus a time varying part, and governed by the equation of state:
\begin{equation}
   \rho =  {\bar \rho} + \rho' = {\bar \rho} (1- \alpha T' - \alpha_\xi \xi'), \label{eq:boussinesq_approximation}
\end{equation}
\noindent where $\alpha_\xi$ is the coefficient of compositional expansion.

Both inner and outer boundaries are stress-free, meaning that viscous stresses vanish at these boundaries as appropriate for the nearly inviscid interiors of giant planets.
The values of temperature and composition are fixed at both boundaries, with $T_i=\xi_i=1$ and/or $T_o=\xi_o=0$, which allows both of these thermodynamic properties to drive convection.

As shown in Table~\ref{tab:derived_parameters}, the Ekman numbers in our models are set to $Ek=10^{-3}$ (cases T1, D, and C1) and $Ek=3 \times 10^{-4}$ (cases T2 and C2); while these values are large compared to planetary estimates, they reflect our available computational resources given the strong degree of supercriticality and this assumption will be revisited in Section \ref{sec:discussion}.
In thermal convection cases, the Prandtl number is chosen to be equal to one following \cite{french2019viscosity} who predict a range of Prandtl numbers between $Pr \sim 0.1 - 10$ for ice giant interiors, with most values being roughly equal to 1.
In geophysical and planetary environments, the ratio of heat and mass diffusivity can be very small \citep[e.g.][]{monville2019rotating}, but we choose $Sc=10$ so that the diffusivities differ by at least one order of magnitude while keeping in mind the limitations of computational resources.
Finally, we choose a value for $\chi = 0.35$, which is consistent with the models of \cite{soderlund2013turbulent} and motivated by the transition to chemically active phases near $\sim 100$ GPa (e.g., $\chi \approx 0.35$ for a uniform density of $\bar \rho \approx 1000$ kg m$^{-3}$).

The Rayleigh numbers were chosen so that our models would have convective turbulence that is three-dimensional with limited vertical stiffness as argued in Section 2. This is achieved by considering the most extreme scaling for the regime transition between the unbalanced boundary layer--non-rotating heat transfer regime identified by \cite{cheng2018heuristic}: $Ra_{UNR} \sim Ek^{-2} Pr$, which can be rewritten in terms of the convective Rossby number:
$Ro_C \sim 1$.
This parameter represents the ratio of buoyancy to Coriolis forces:

\begin{equation} \label{eq:convective_rossby}
	\begin{split}
		Ro_C^2 & = \frac{\alpha \Delta T g + \alpha_\xi \Delta \xi g}{D \Omega^2}
	 	= Ek^2 \left( \frac{Ra}{Pr} + \frac{Ra_\xi}{Sc} \right),
	\end{split}
\end{equation}
\noindent Convection is weakly constrained by rotation when $Ro_C \gg 1$ and the dynamics are rotationally dominated when $Ro_C \ll 1$ \citep{gastine2016scaling}.
In our cases, we will choose $Ro_C$ to be equal to $\sqrt{2}$ following \cite{soderlund2013turbulent}.
In the double diffusive case, the Rayleigh numbers are chosen so that the coefficient of the thermal and compositional buoyancy term in Equation~\ref{eq:navier_stokes} are equal.
The resulting Rayleigh numbers range in supercriticality from $Ra/Ra_C \sim 20 $ for case T1 to nearly $Ra_\xi/Ra_C \sim 400$ for case C2 (Table~\ref{tab:derived_parameters}), where $Ra_C$ are the critical Rayleigh numbers for the onset of convection calculated using the Spherical INertia-Gravity Eigenmodes (SINGE)\footnote{https://gricad-gitlab.univ-grenoble-alpes.fr/Geodynamo/Singe} code \citep{vidal2015quasi}.

We next consider these dimensionless input parameters in terms of the associated dimensional properties. The outer radius, shell thickness, and rotation rate are fixed to $r_o \approx 2.5 \times 10^7$ m, $D \approx 1.6 \times 10^7$ m and $\Omega \approx 10^{-4}$ s$^{-1}$ in all models.
Given the choices of Ekman number, this implies a kinematic viscosity of $\nu = [3 \times 10^7,8 \times 10^6]$ m$^2$ s$^{-1}$ for $Ek = [10^{-3}, 3\times 10^{-4}]$.
Setting $Pr=1$ implies that the thermal diffusivity is equal to the viscosity: $\kappa = [3 \times 10^7, 8 \times 10^6]$ m$^2$ s$^{-1}$ for $Ek = [10^{-3}, 3\times 10^{-4}]$.
Conversely, $Sc=10$ implies that the mass diffusivity is an order of magnitude smaller: $\kappa_\xi=[2.6 \times 10^6, 8 \times 10^5]$ m$^2$ s$^{-1}$ for $Ek = [10^{-3}, 3\times 10^{-4}]$.
If we assume that the fluid in the interior is water, then the thermal expansion coefficient is on the order of $\alpha \sim 10^{-4}$ K$^{-1}$ \citep{mazevet2019ab} and the temperature contrast is approximately 390 K in both thermal cases and 195 K in the double diffusive case.
In order to estimate the composition contrast, we first calculate the density of a mixture of hydrogen and water using the linear mixing approximation \citep{bethkenhagen2017linear}.
Interpreting $\xi$ as the mass fraction of hydrogen, the density of the mixture is given by

\begin{equation}
    \frac{1}{\rho_M} = \frac{\xi}{\rho_H} + \frac{1 - \xi}{\rho_W},
\end{equation}

\noindent where $\rho_M$ is the density of the mixture, $\rho_H$ is the density of hydrogen and $\rho_W$ is the density of water, taken at identical pressure and temperature conditions.
For small values of $\xi$, we can approximate the expansion coefficient as

\begin{equation}
    \alpha_\xi = - \frac{1}{\rho} \frac{\partial \rho}{\partial \xi} \approx \left(\frac{\rho_W}{\rho_H} - 1 \right) - \left( \frac{\rho^2_W}{\rho^2_H} - 1 \right) \xi + \mathcal{O}(\xi^2).
\end{equation}

It is reasonable to assume that in the planetary interiors, light elements would infuse slowly into the convective region of the interior.
For this reason, the concentration of the fluid most enriched in light elements near the inner boundary will be sufficiently small that this approximation is applicable and $\Delta \xi \ll 1$.
Based on equations of state for water and hydrogen, we find that typical densities that coexist in hydrogen and water are $\rho_H \sim 600$ kg m$^{-3}$ and $\rho_W \sim 2,500$ kg m$^{-3}$, respectively \citep{chabrier2019new,mazevet2019ab}.
This gives $\alpha_\xi \sim 3.2$, which corresponds to $\Delta \xi \sim  0.02$ kg kg$^{-1}$ for both compositional cases and $\Delta \xi \sim 0.01$ kg kg$^{-1}$ for the double diffusive case.

Many of these physical values (e.g., diffusivities, thermal and compositional anomalies) are necessarily much larger than those expected for the ice giants. Considering kinematic viscosity as an example, $\nu$ is more than 10 orders of magnitude too large, although exploiting a larger turbulent eddy viscosity interpretation would reduce this disparity. 
However, given the limited resolution of Uranus and Neptune observations, we assume kinematic similarity exists between the large-scale flows of ice giants interiors and our models.
It is also broadly instructive to understand how varying the source of convective driving effects the dynamics in rotating, spherical systems, and how these differences may manifest in nature.

\subsection{Numerical method}

Equations \ref{eq:navier_stokes} -- \ref{eq:mass} are solved numerically using the pseudospectral fluid dynamics code MagIC\footnote{magic-sph.github.io} compiled using the SHTns library\footnote{https://bitbucket.org/nschaeff/shtns/src/master/} of spherical harmonic transforms to improve performance \citep{wicht2002inner,breuer2010thermochemically,schaeffer2013efficient}.
MagIC takes advantage of Equation \ref{eq:boussinesq} by decomposing the velocity into two scalar potentials: a poloidal potential and toroidal potential.
Non-linear terms of the model equations are calculated on a grid in spherical coordinates $(r,\theta, \phi)$ of $N_R \times N_\phi/2 \times N_\phi$, where $N_R$ is the number of radial points and $N_\phi$ is the number of longitudinal grid points.
Radial grid points are distributed over Chebyshev nodes, which allows for better resolution near the inner and outer boundary in addition to allowing for efficient transforms between the grid and spectral representation.
Latitudinal grid points placed on Gauss-Legendre quadrature nodes and longitudinal grid points are distributed uniformly.
All quantities have a spectral representation, with angular dependence in terms of spherical harmonics, $Y^m_n (\theta, \phi)$ and radial dependence in terms of Chebyshev polynomials, with the spectral representation being truncated at a maximum spherical harmonic degree $n_{max}$ and Chebyshev degree $N_R-2$.
Table \ref{tab:derived_parameters} contains the resolutions used each case.
The higher Ekman number cases were initialized as solidly rotating bodies ($\textbf{u}=0$), with perturbations to the temperature and/or composition seeded as white noise.
The lower Ekman number cases were initialized using equilibrated states of the corresponding higher Ekman number cases.
Once a statistically consistent state is achieved, the solutions are time-averaged over at least one viscous diffusion time to obtain mean quantity values.

\section{Results \label{sec:results}}

\begin{deluxetable}{cCCCcccccccccCC}
    \tablecaption{Control and time-averaged derived parameters for each case.
    The control parameters are defined in Equations \ref{eq:Ekman} -- \ref{eq:Rayleigh}, \ref{eq:Schmidt}, and \ref{eq:ChemicalRayleigh}. Spatial resolution is given by the number of radial grid points $N_r$ and maximum spectral harmonic degree $n_{max}$.  The duration of time-averaging is given in terms of the viscous dissipation time scale.  Three derived parameters that characterize the velocity field include: the Reynolds number, $Re$,  the local Rossby number $Ro_l$, and characteristic spherical harmonic degree, $\bar n$, defined in Equation~\ref{eq:mean_degree}.  The Nusselt number, $Nu$, and Sherwood number, $Sh$, characterize the thermodynamic fields and are defined in Equation \ref{eq:nusselt} and \ref{eq:sherwood}, respectively.  Coefficients of variation (CV) are provided for the temperature, composition, and the axial component of absolute angular momentum which are calculated using Equation \ref{eq:cov}.
    \label{tab:derived_parameters}}
    \tablehead{
        \colhead{Case} & \colhead{$Ek$} & \colhead{$Ra$} & \colhead{$Ra_\xi$} & \colhead{$N_r$} & \colhead{$n_{max}$} & \colhead{Duration} & \colhead{$Re$} & \colhead{$Ro_\ell$} &  \colhead{$\bar n$} & \colhead{$Nu$} & \colhead{Snapshots} & \colhead{$CV_T$} & \colhead{$CV_{h_z}$} \\
        \colhead{} & \colhead{$\times 10^{-3}$} & \colhead{} & \colhead{} & \colhead{} & \colhead{} & \colhead{} & \colhead{} & \colhead{} &  \colhead{} & \colhead{($Sh$)} & & \colhead{($CV_\xi$)} & }
    \startdata
    T1 & 1 & 2\times 10^6 & \nodata & 61 & 128 & 1.50 & 335 & 0.39 & 3.64 & 7.0 & 75 & 0.91 & 0.52 \\
     & & (21 Ra_C) & & & & & & & & \nodata & & \nodata & & \\
    \hline
    D & 1 & 1\times 10^6 & 1\times 10^7 & 61 & 134 & 1.20 & 263 & 0.31 & 3.72 & 6.1 & 60 & 0.85 & 0.56 \\
     & & & & & & & & & & (20.9) & & (0.44) & & \\
    \hline
    C1 & 1 & \nodata & 2\times 10^7 & 101 & 192 & 2.25 & 95 & 0.26 & 8.75 & \nodata & 123 & \nodata & 0.66 \\
    & & & (134 Ra_C) & & & & & & & (16.5) & & (0.48) & & \\
    \hline
    T2 & 0.3 & 2.22\times 10^7 & \nodata & 65 & 192 & 1.22 & 1096 & 0.43 & 4.11 & 12.9 & 121 & 0.79 & 0.46 \\
    & & (66 Ra_C) & & & & & & & & \nodata & & \nodata & & \\
    \hline
    C2 & 0.3 & \nodata & 2.22\times 10^8 & 161 & 213 & 1.02 & 615 & 0.25 & 4.21 & \nodata & 109 & \nodata & 0.55 \\
    & & & (376 Ra_C) & & & & & & & (37.1) & & (0.34) & & \\
    \enddata
\end{deluxetable}

\subsection{Velocity field \label{sec:velocity_results}}

\begin{figure}
    \fig{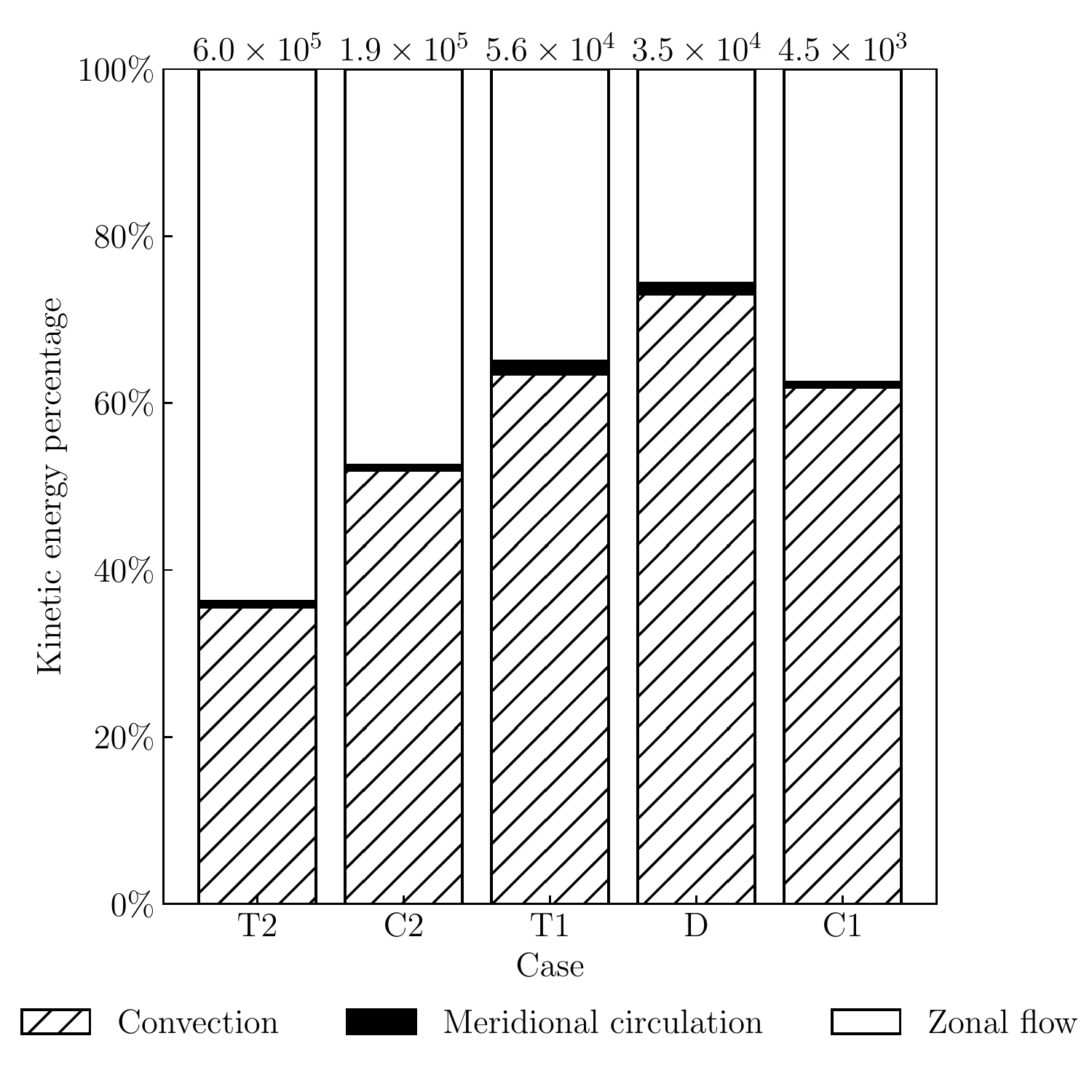}{.5\linewidth}{}
	\caption{Bar plot of the percentage of kinetic energy that is represented by axisymmetric zonal flows, axisymmetric meridional circulations, and the non-axisymmetric convection.  The time-averaged value of the total non-dimensional kinetic energy for each case is given above the bar corresponding to that case. \label{fig:ekin_bar}}
\end{figure}

\begin{figure}
    \fig{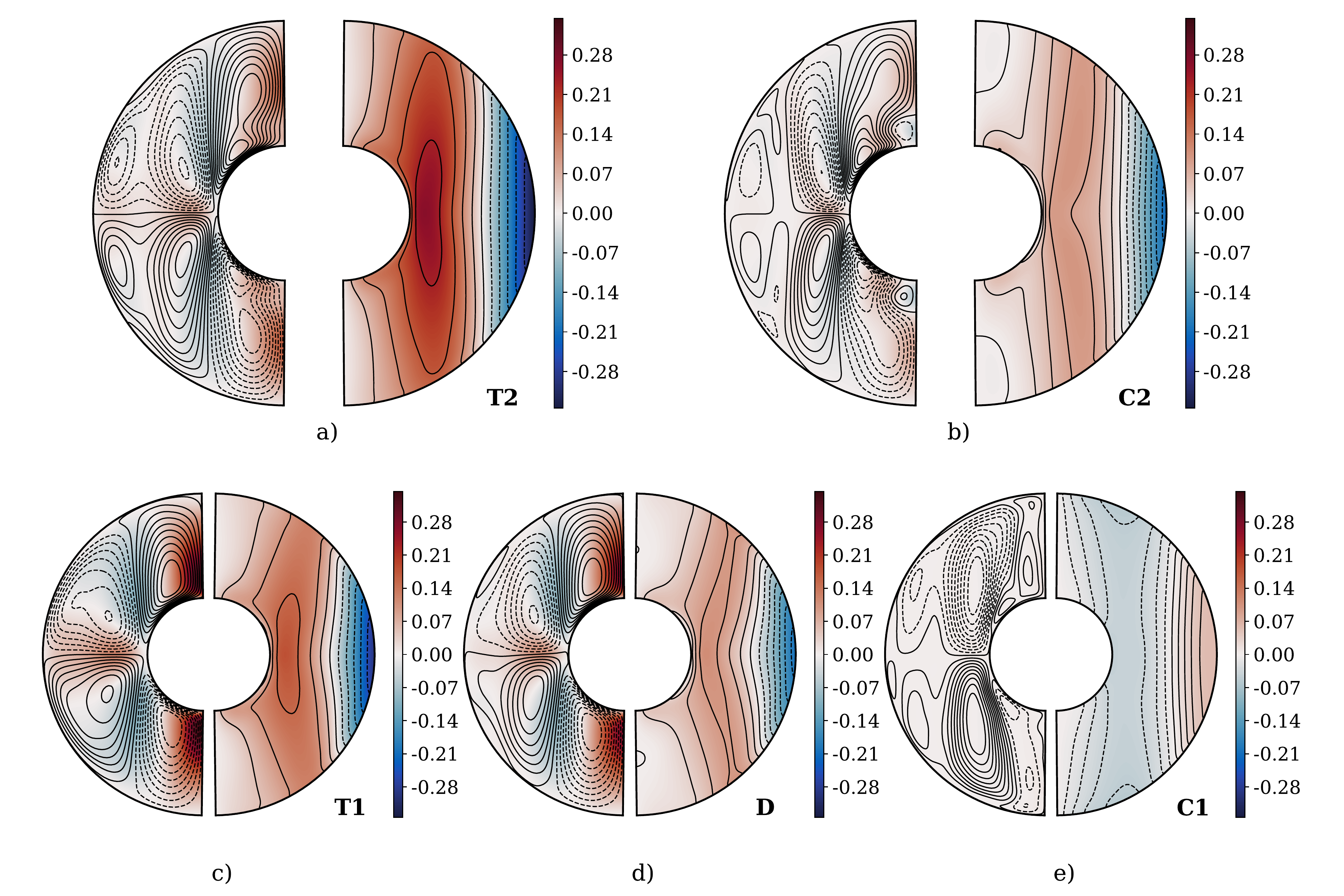}{\linewidth}{}
	\caption{Meridional slices of time- and azimuthally-averaged meridional circulation (left panels) and zonal velocity (right panels) for each case.  Lines of meridional circulation are superimposed on the radial component of velocity.  Both radial and zonal velocity are expressed in terms of the Rossby number, $Ro = u/r_o \Omega$.  In plots of the zonal velocity, red regions represent flow in a prograde direction, and blue represents flow in a retrograde direction.  In plots of the radial velocity, red regions represent outward flow, and blue regions represent inward flow.  Lines in the left panels are lines of circulation, where solid (dashed) lines represent counterclockwise (clockwise) rotation.  Lines in the right panels are level curves of constant zonal velocity, where solid (dashed) lines represent prograde (retrograde) flows.}
	\label{fig:zonal_flow}
\end{figure}

Figure~\ref{fig:ekin_bar} shows the relative percentages of the kinetic energy contained in different components of the flow. 
The axisymmetric poloidal and axisymmetric toroidal components of the kinetic energy correspond to the zonal flows and meridional circulations, respectively, while the non-axisymmetric components correspond to convective motions.
The zonal flows tend to dominate the dynamics of these systems in the lower $Ek$ cases (T2 and C2), while non-symmetric components dominate in the higher $Ek$ cases (T1, D and, C1).
The zonal flows, in addition to being axisymmetric, are mostly equatorially symmetric (see Figure~\ref{fig:zonal_flow}).
In all cases, meridional circulation represents less than 2\% of the total kinetic energy, meaning that these systems may be interpreted as consisting primarily of zonal flows and convection.
Comparing T2 and C2, while the absolute kinetic energy associated with convection is larger in T2 ($2.1 \times 10^5$ in T1 compared to $9.8 \times 10^4$ in C2) , the fractional contribution of convection increases with the decrease in diffusivity.
However, comparing T1 and C1, the same changes do not occur, with similar percentages for convective and zonal flow kinetic energy relative to the total despite different absolute kinetic energies.
The zonal flow thus tends to decrease in total fraction of kinetic energy moving from T2 and C2, and moving from T1 to D, indicating that the amplitude of the zonal winds is influenced by the thermodynamics of the system.
This trend does not continue when moving from D to C1, which behaves differently from the others in several regards as described further below.

The characteristic spherical harmonic degree of the flow can be calculated from the kinetic energy spectra following \cite{christensen2006scaling}:
\begin{equation}
    \bar{n}  = \frac{\sum_n n \textbf{u}_n^2}{\sum_n \textbf{u}^2_n} \label{eq:mean_degree},
\end{equation}
\noindent where $\mathbf{u}_n$ is the velocity associated with all spherical harmonics of degree $n$.
With the notable exception of C1, the cases tend to have similar $\bar{n}$ values for a given Ekman number, differing by about 2\% between C2 and T2 and T1 and D. Case C1, however, is a clear outlier with $\bar{n}$ more than twice that of T1 and D.
When the Rossby number (defined in Equation~\ref{eq:rossby}) is calculated on the basis of the length scale $D \pi/\bar n$ as described in Equation \ref{eq:mean_degree}, it is known as the local Rossby number, and better reflects the ratio of inertia to the Coriolis force in the force balance of the system \citep{christensen2006scaling}. 
The local Rossby numbers, tabulated in Table~\ref{tab:derived_parameters}, span the range $0.25 \leq Ro_\ell \leq 0.43$ and are roughly similar in cases with the same mode of convective driving.
The largest values occur in T2 and T1, while the smallest values appear in C2 and C1, with D being intermediate between the thermal and compositional cases.
The Coriolis force is thus expected to exceed inertia on convective length scales, but not by a significant amount.

Figure~\ref{fig:zonal_flow} shows the time- and azimuthally-averaged meridional circulations and zonal flows. 
All cases have three-banded zonal flows, with four of the five cases having qualitatively ice-giant-like flows along the outer boundary, i.e. a retrograde jet at low latitudes and prograde jets at higher latitudes.
As the Ekman number is decreased for the thermal convection cases (T1 $\rightarrow$ T2), equivalent to decreasing the fluid viscosity, the retrograde flows intensify by 17\%.
In contrast, the zonal jets reverse directions between the two compositional convection cases (C1 $\rightarrow$ C2), with prograde equatorial flows and retrograde higher latitude jets developing in the case of C1.
While the zonal flows tend to have relatively small variations along the axial $\hat{\mathbf{z}}$ direction at large cylindrical distances, more significant vertical variations develop near and within the tangent cylinder. 
As convective driving is varied (T $\rightarrow$ C), there is a decrease in the flow intensity across the system, by 9\% between T2 and C2 and by 120\% between T1 and C1 (which is made especially large given the reversal).
There is also a change in the characteristics of vertical variations of these flows, with a particularly intense prograde belt in the deep interior of T1 and T2, which does not appear in D or C2.
In C1, in both the polar region (within the tangent cylinder) and far from the axis of rotation near the outer boundary, the zonal flow is constant along nearly vertical surfaces.

Meridional circulations contain comparable fractions of kinetic energy across all of the models and similarly all feature, to some extent, fluid upwelling from the lower boundary near the equator (Figure~\ref{fig:zonal_flow}). However, the circulation patterns also have notable differences between them.
In each case, mid-latitude circulation cells are present, but the number and characteristics of the circulation cells change as both Ekman number and convective driving are varied.
T2 has two circulation cells outside the tangent cylinder in each hemisphere, whereas single cells develop with faster radial velocities appear in the case of T1.
Inside the tangent cylinder, additional circulation cells appear near the inner boundary as the Ekman number is decreased from T1 to T2. 
Cases T1 and D behave quite similarly, although flows are weaker in the latter.
Compared with T2, C2 has two additional circulation cells that form outside of the tangent cylinder at the mid latitudes.
Decreasing the Ekman number in the compositional convection cases (C1 $\rightarrow$ C2, the extent of the polar cell increases and an additional cell appears near the axis of rotation.
In varying the convective driving from D to C1, this polar cell diminishes in its extent and the mid-latitude cell crosses into the tangent cylinder.

\begin{figure}
    \fig{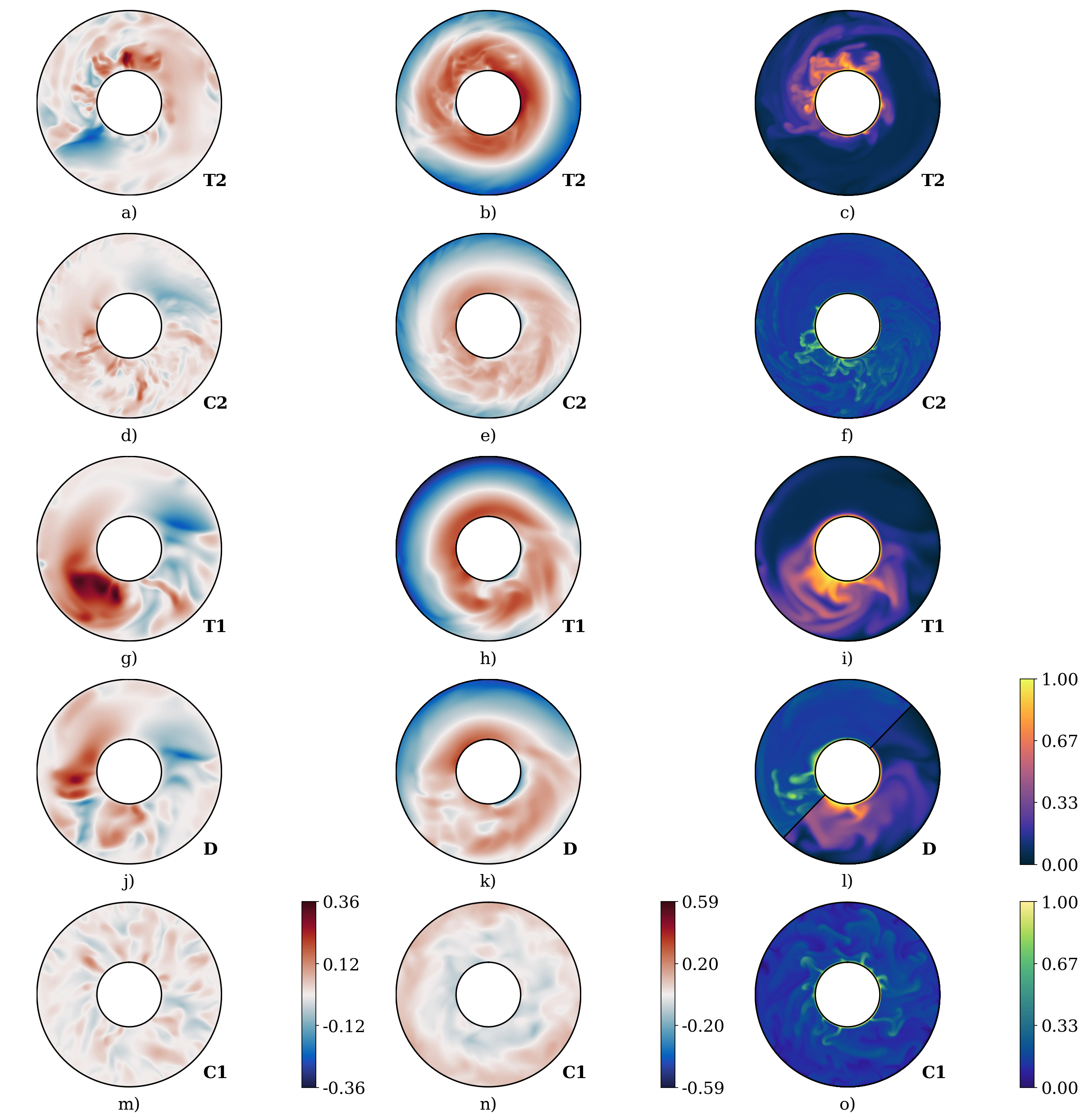}{\linewidth}{}
    \caption{Equatorial slices of radial velocity (left), zonal velocity (middle), and temperature/composition (right) for all cases at snapshots in time. For radial velocity, red (blue) indicates outward (inward) motion.  For zonal velocity, red (blue) indicates prograde (retrograde) motion.  All velocities are given in $Ro=u/r_o \Omega $ units. For temperature (composition), bright colors represent hotter (higher concentration) fluid, while darker colors represent colder (lower concentration) fluid. For a given field, the color bar in each sub-Figure~displays the same range of values. Because the double diffusive case has both temperature and composition fields, we represent this in l) by splitting the equatorial plane into two hemispheres, representing the composition in the top-left half, and the temperature in the bottom-right.  The field is divided to show part of both the active and passive hemisphere for each field.} 
    \label{fig:equatorial_slice}
\end{figure}

Figure~\ref{fig:equatorial_slice} shows slices in the equatorial plane of radial and zonal velocity and thermodynamic fields.
Similar trends appear in the intensity of velocities as those seen in Figure~\ref{fig:zonal_flow}.
The flow becomes more disorganized in the lower Ekman number cases, where there are more small scale features in the components of the velocity field.
In the cases with retrograde equatorial winds, one hemisphere is thermally/compositionally active, while the other is quiet, which leads to differences in the structure of the velocity field. 
Convective motions associated with the radial velocity are more intense and show more small scale features in the active hemisphere.
As the fluid flows from the quiet to the active hemisphere, a region of subsumption forms, causing the colder (light element depleted) fluid to flow back toward the inner boundary.
Along with this, the zonal flows in the active hemisphere become more disturbed, with a smoother transition from prograde to retrograde flow from the inner to outer boundary in the quiet hemisphere.
In the lower Ekman cases, the equatorial jet wraps more completely around the circumference of the system.

In the case of C1, the properties of the flow are significantly different.
Prograde zonal flows appear to encircle the entire system and the system no longer appears to be divided into an active and quiet hemisphere, with radial upwelling and downwelling occurring across all longitudes.
For this reason, the zonal flows appear to be disturbed at all longitudes as well.

\begin{figure}
    \gridline{
        \fig{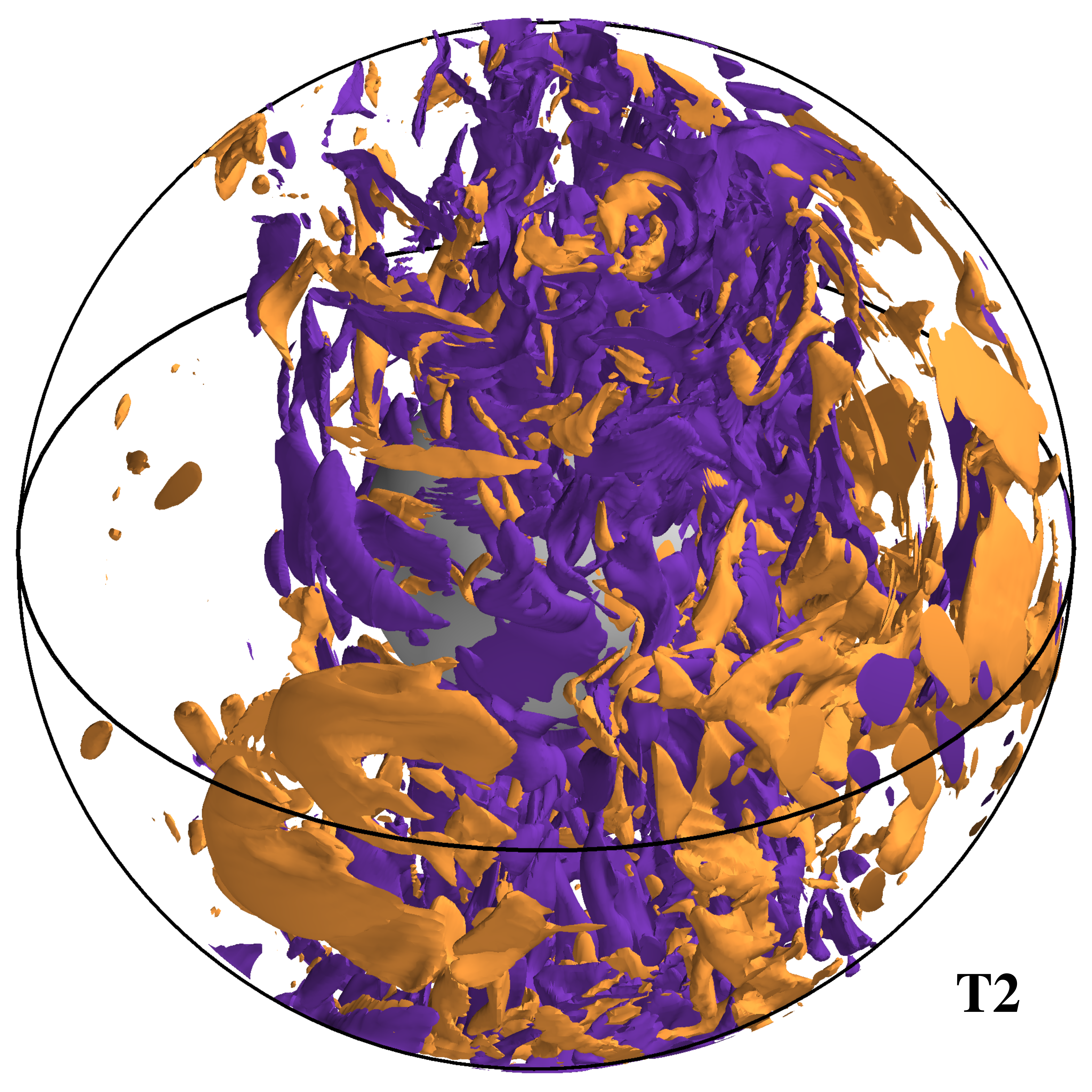}{.4\linewidth}{\large{a)}}
       \fig{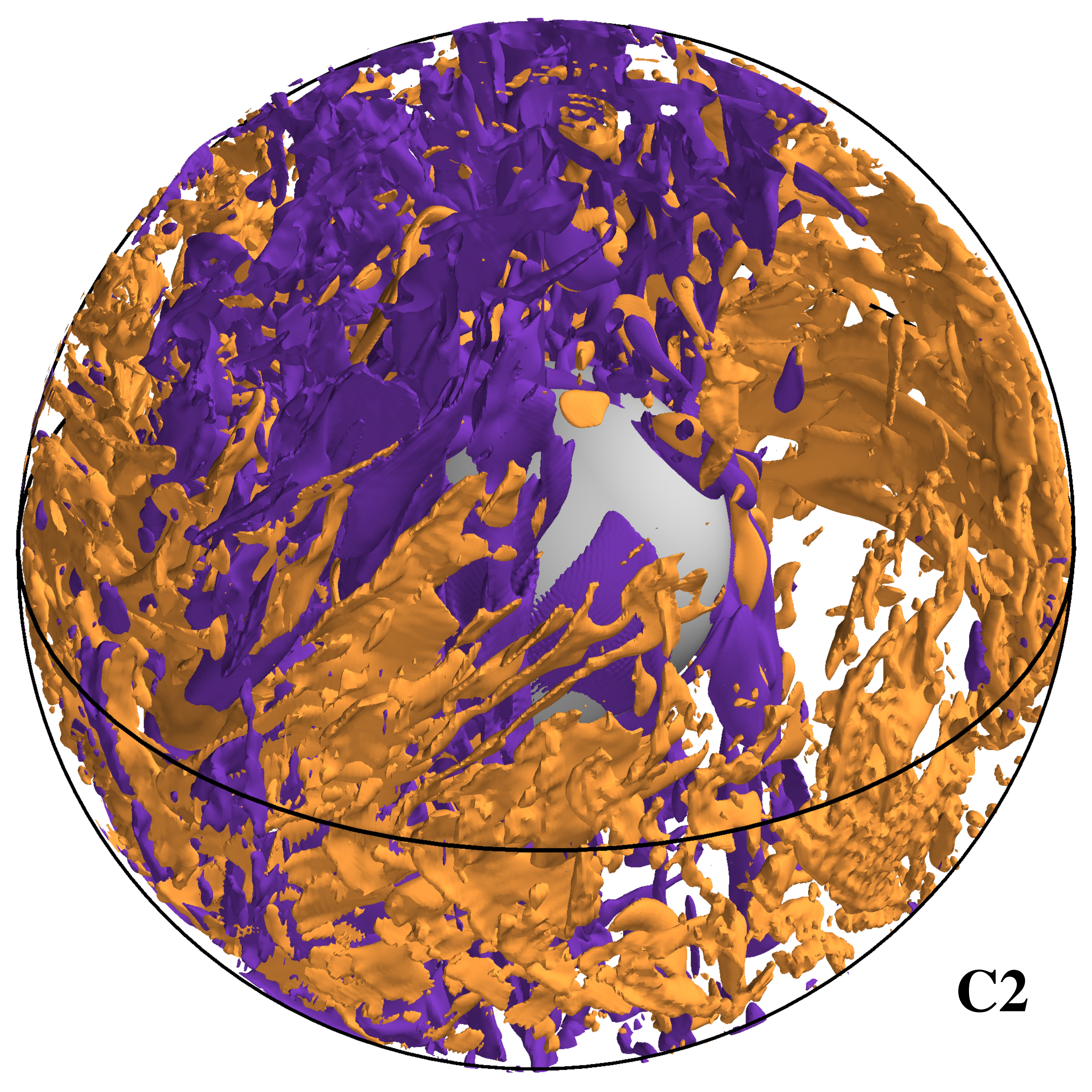}{.4\linewidth}{\large{b)}}}
    \gridline{
        \fig{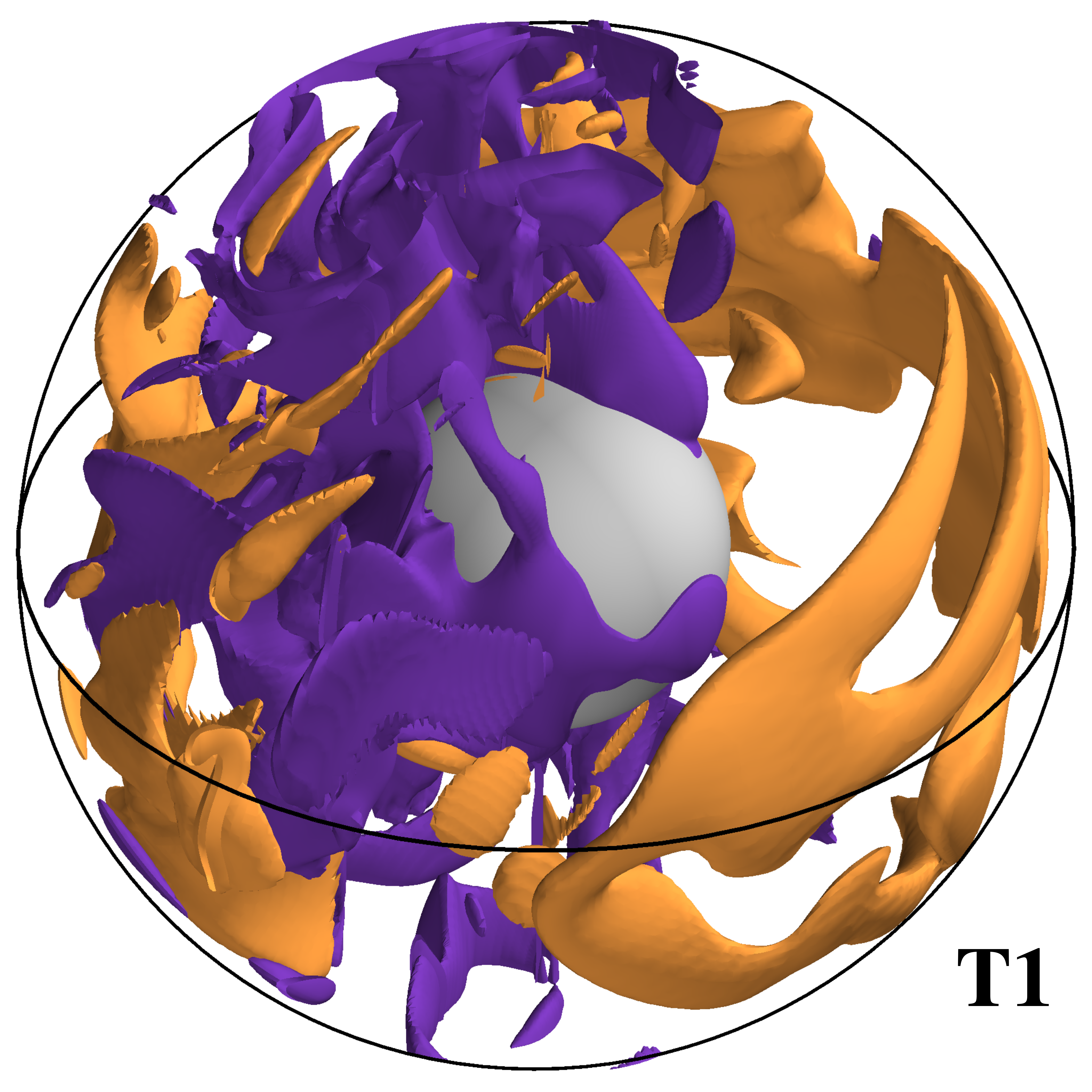}{0.2667\linewidth}{\large{c)}}
        \fig{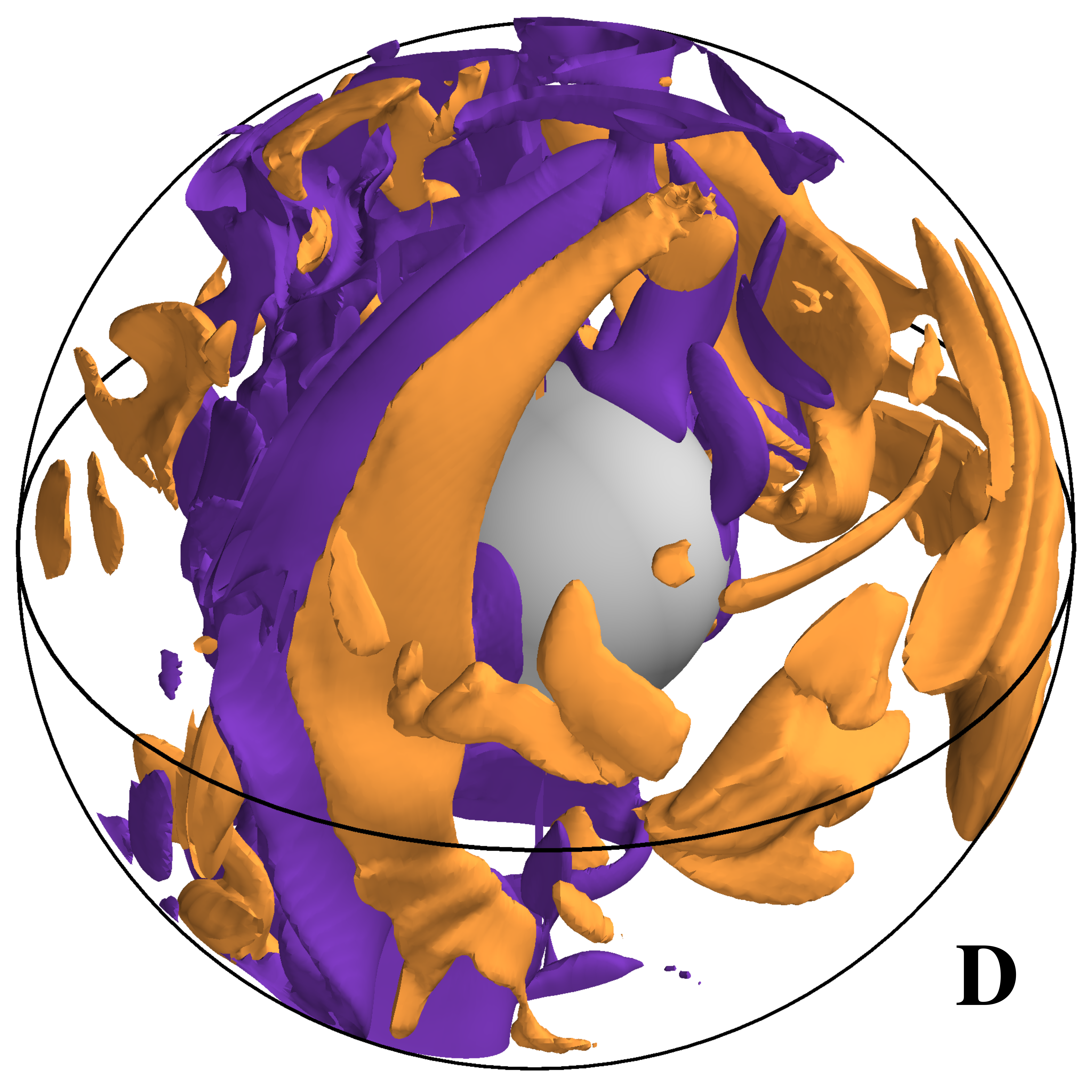}{0.2667\linewidth}{\large{d)}}
        \fig{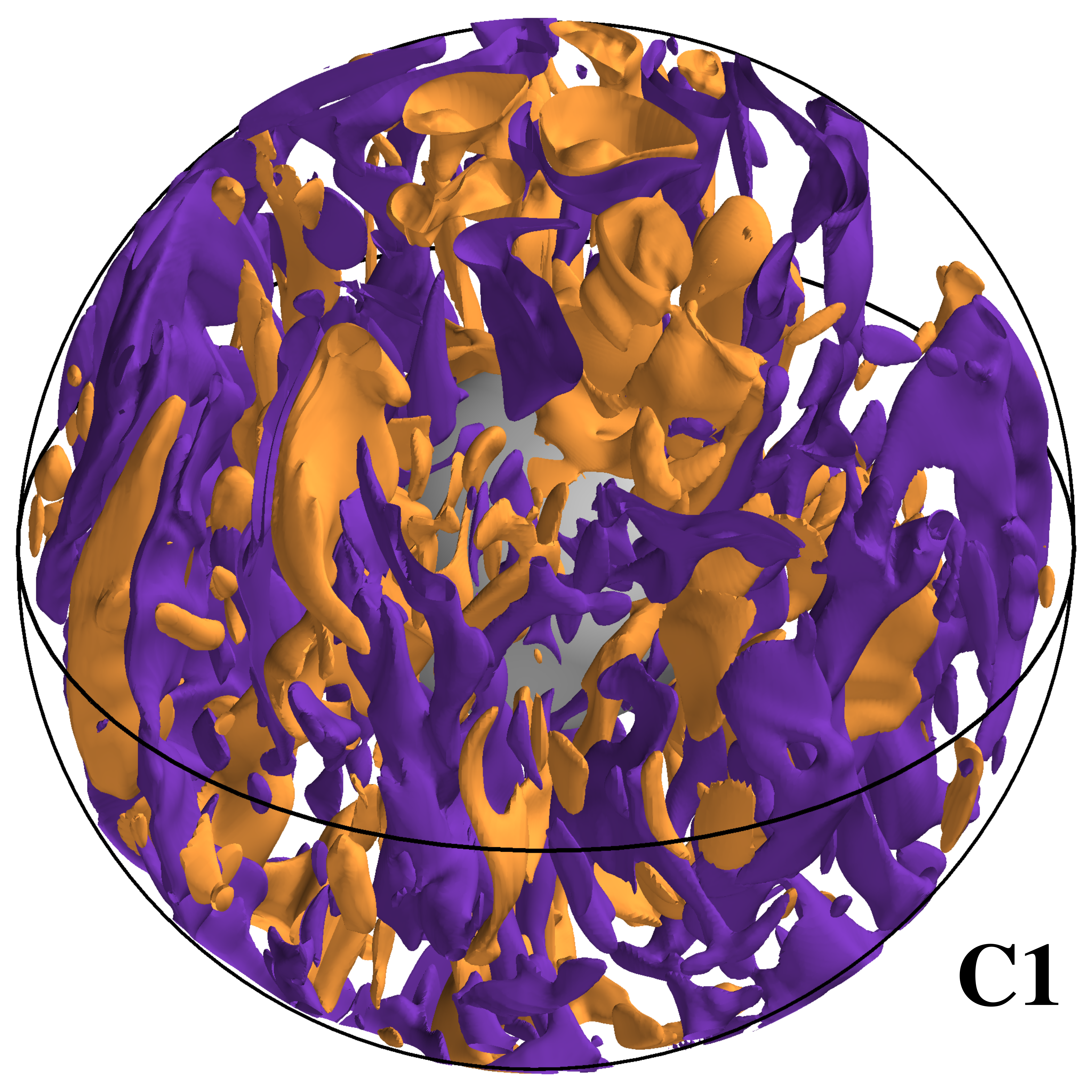}{0.2667\linewidth}{\large{e)}}}
    \caption{Isosurfaces of the axial component of the vorticity, $\omega_z=\nabla \times {\bf u} \cdot {\hat {\bf z}}$, for each case.  Orange (purple) isosurfaces represent anticyclonic (cyclonic) vorticity.  Values of the isosurfaces are a) $|\omega_z|=1.14\times10^4$ for T2, b) $|\omega_z|=6.12\times10^3$ for C2, c) $|\omega_z|=2.28\times10^3$ for T1, d) $|\omega_z|=1.81\times10^4$ for D, and e) $|\omega_z|=1.28\times10^3$ for C1.  The grey sphere in the interior represents the inner boundary, and the black circles indicate the equator and a great circle facing the viewer on the outer boundary.}
    \label{fig:3d_vorticity}
\end{figure}

Figure~\ref{fig:3d_vorticity} plots typical three-dimensional isosurfaces of the axial component of vorticity for each of the cases, illustrating the three-dimensional nature of the velocity fields that has developed.
When rotation dominates the system, the Taylor-Proudman theorem dictates that the properties of the fluid should be organized into columnar structures (i.e. invariant in $z$).
In all cases except for C1, the flows that develop are not strongly organized by rotation as constructed a priori.
The distribution of isosurfaces is similar to the features that develop in the temperature and composition fields in Figure~\ref{fig:equatorial_slice}c, f, i, l, and o.
That is, in cases with active and quiet hemispheres, the vorticity is mostly concentrated in the active hemisphere.
Again, C1 appears to be different, with features distributed more evenly across all longitudes and that tend to exhibit some alignment in the axial direction.

\subsection{Temperature and composition}

\begin{figure}
    \fig{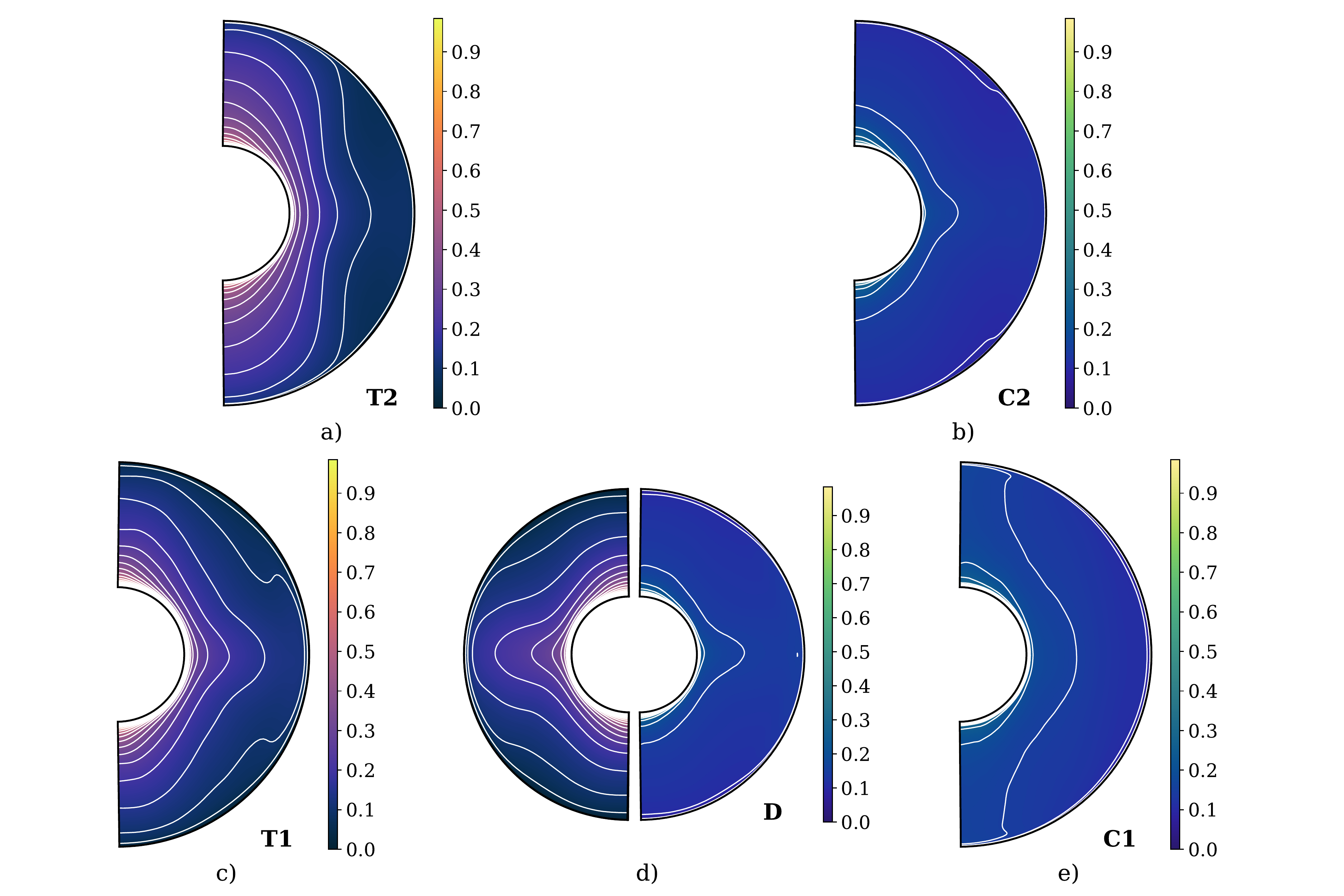}{\linewidth}{}
    \caption{Meridional slices of time-averaged and azimuthally-averaged temperature/composition for all cases. For temperature (composition), bright colors represent hotter (higher concentration) fluid, while darker colors represent colder (lower concentration) fluid.  \label{fig:temperature_slices}}
\end{figure}

Figure~\ref{fig:temperature_slices} plots meridional slices of time- and azimuthally-averaged temperature and/or composition for each case.
Heat and mass are transported by advection and diffusion as described in Equations \ref{eq:heat} and \ref{eq:mass}.
Temperature (composition) profiles in which heat (light elements) is transported only by diffusion can be predicted using Laplace's equation, which will have a particular gradient and the property that these solutions are radially symmetric.
In these cases, the presence of turbulent motion will act to improve the ability of heat and mass to become more well mixed in the bulk fluid, which would decrease the steepness of the interior temperature and/or concentration gradient compared to that predicted by diffusion alone.
When controlled for viscosity, greater diffusivity of heat or mass leads to a steeper gradient near the boundaries, but a shallower gradient throughout the bulk of the fluid.
When controlled for convective driving, though the effect is less pronounced, the gradient in the bulk of the fluid becomes shallower.
Because this is equivalent to reducing the diffusivity of the thermodynamic field, it effectively causes the field to adhere less closely to the diffusive solution.
In addition, the latitudinal dependence of this gradient reflects the differing dynamics within the convective region.
The greatest degree of mixing occurs along the equatorial plane and mid-latitudes, whereas the gradient tends to become steeper near the polar regions.
This results from the fluid motions in the equatorial plane, which tend to be more intense that those that form within the polar regions.
Each case has equatorial upwellings, with multiple cells forming outside the tangent cylinder, which will mix the fluid.
Most of the poloidal kinetic energy is equatorially symmetric, which implies a local extremum along the equatorial plane.

\begin{figure}
    \fig{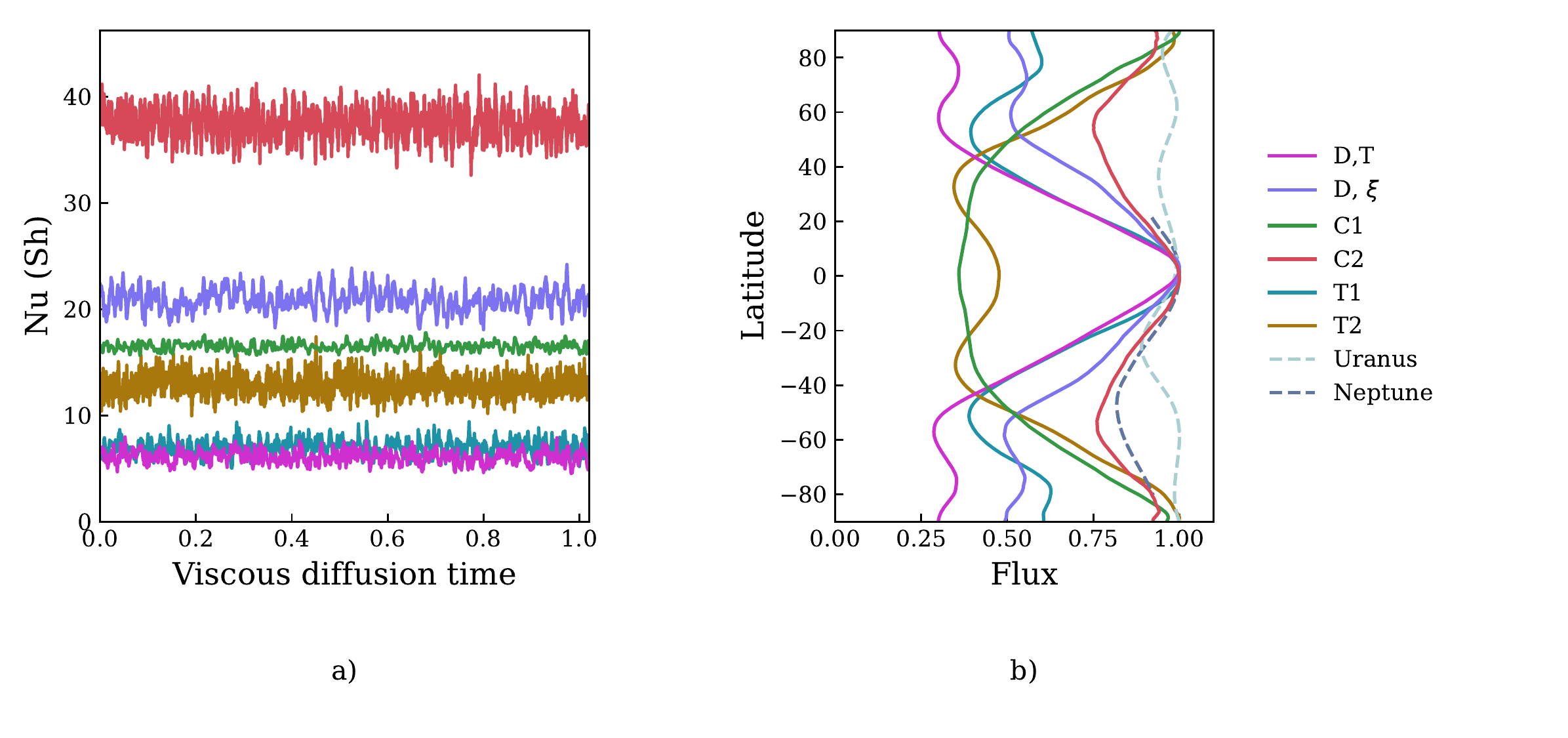}{\linewidth}{}
    \caption{Thermal (compositional) properties of each case, presented as a) the Nusselt (Sherwood) number as a function of time and b) the flux in terms time- and azimuthally-averaged Nusselt (Sherwood) number as a function of latitude for both our simulations and for Uranus and Neptune \citep{pearl1990albedo,pearl1991albedo,soderlund2013turbulent}.  In a), the horizontal axis uses the non-dimensional time units and all time-series are truncated to fit within the same window.  In b), the radial derivatives of temperature and/or composition are normalized by their maximum values.}
    \label{fig:flux_latitude}
\end{figure}

The Nusselt number, $Nu$, and its compositional analog, the Sherwood number, $Sh$, are the non-dimensional heat and compositional gradients across the outer boundary, respectively \citep{incropera2007fundamentals}.
The Sherwood number is calculated similarly to the Nusselt number in Equation \ref{eq:nusselt}:
\begin{equation}
	Sh = \frac{\dot \xi D}{\rho \kappa_\xi} \label{eq:sherwood},
\end{equation}
\noindent where $\dot \xi$ is the mass flux of light elements, measured in kg m$^{-2}$ s$^{-1}$.
Figure~\ref{fig:flux_latitude}a contains time series of these quantities, and time-averaged values are also tabulated in Table \ref{tab:derived_parameters}.
They tend to be consistent across the duration of simulations, with most cases being distinct from one another.
As the Ekman number is decreased for a fixed $Ro_c$, the Nusselt and Sherwood numbers both increase due to the stronger convective motions that result from the lower viscosity and stronger driving force.
Comparing T1 to C1 and T2 to C2, there is also an increase in the convective transport, as reducing the Prandtl or Schmidt number decreases the relative importance of diffusive heat or mass transport.
While D has a comparable, yet smaller, Nusselt number relative to T1, it has a larger Sherwood number than C1; this is likely a consequence of the faster flows that develop in D compared to C1, resulting in more convective transport.
The range of Nusselt and Sherwood numbers does not have a distinct trend with respect to the convective driving; however, because compositional fields are associated with a higher flux than the temperature fields, the relative variability decreases with the decrease in thermal (mass) diffusivity.

Figure~\ref{fig:flux_latitude}b plots heat and mass flux as a function of latitude.
Local maxima occur near the equator in all cases except C1.
The cases show two trends in the flux near the poles.
T2, C1, and C2 have polar maxima, while D (both temperature and composition) and T1 do not, though additional local maxima do appear at higher latitudes.
In all cases except C1 that has a global minimum at the equator, local minima appear at mid- to high- latitudes.
Because these measure the relative importance of convective heat and mass transfer, they depend on the structure of the velocity field.
The effect of decreasing viscosity tends to be opposite when controlled for convective driving.
Comparing T2 to T1, the peak at the equator is suppressed, with the global maximum appearing at the poles.
The opposite is true in comparing C2 to C1, where the variation with latitude is smaller in C2 as peak near the equator has been raised in comparison to C1.
Similarly, the effects of varying convective driving do not have consistent effects, when controlled for viscosity.
Furthermore, we find that the contrast in heat (mass) flux as a function of latitude becomes smaller (i.e. the curve is flatter) as P\'eclet number increases.

\section{Analysis \label{sec:analysis}}

\subsection{Angular momentum}

Properties of the velocity field are related to angular momentum, and conservation of absolute angular momentum is important to determining the dynamics of the system.
Absolute angular momentum is the total angular momentum in the non-rotating reference frame resulting from solid body rotation and differential rotation of the fluid.
Radial transport of angular momentum outward by individual fluid parcels can result in decreasing zonal flow, if angular momentum for the parcel is conserved, in the same way that a spinning skater extending their arms is able to reduce their rate of rotation \citep{gilman1979angular}.
Retrograde equatorial winds are hypothesized to form when absolute momentum becomes mixed, either completely or partially, in the fluid contained within a spherical shell \citep{aurnou2007effects, gastine2013zonal}.
Because the system rotates around the ${\bf \hat{z}}$-axis, we focus on the component of absolute angular momentum in this direction, $h_z$.
We normalize this quantity by the value of solid-body angular momentum at a distance of $r_o$ from the axis of rotation:
\begin{equation}
    h_z = \frac{s u_\phi + \Omega s^2}{\Omega r_o^2} = (1 - \chi) r \sin \theta Ro_\phi + (1 - \chi)^2 r^2 \sin^2 \theta, \label{eq:angular_momentum}
\end{equation}
\noindent where $s=r \sin \theta$ is the perpendicular distance from the axis and $Ro_\phi$ is the Rossby number of the zonal flow.

\begin{figure}
    \fig{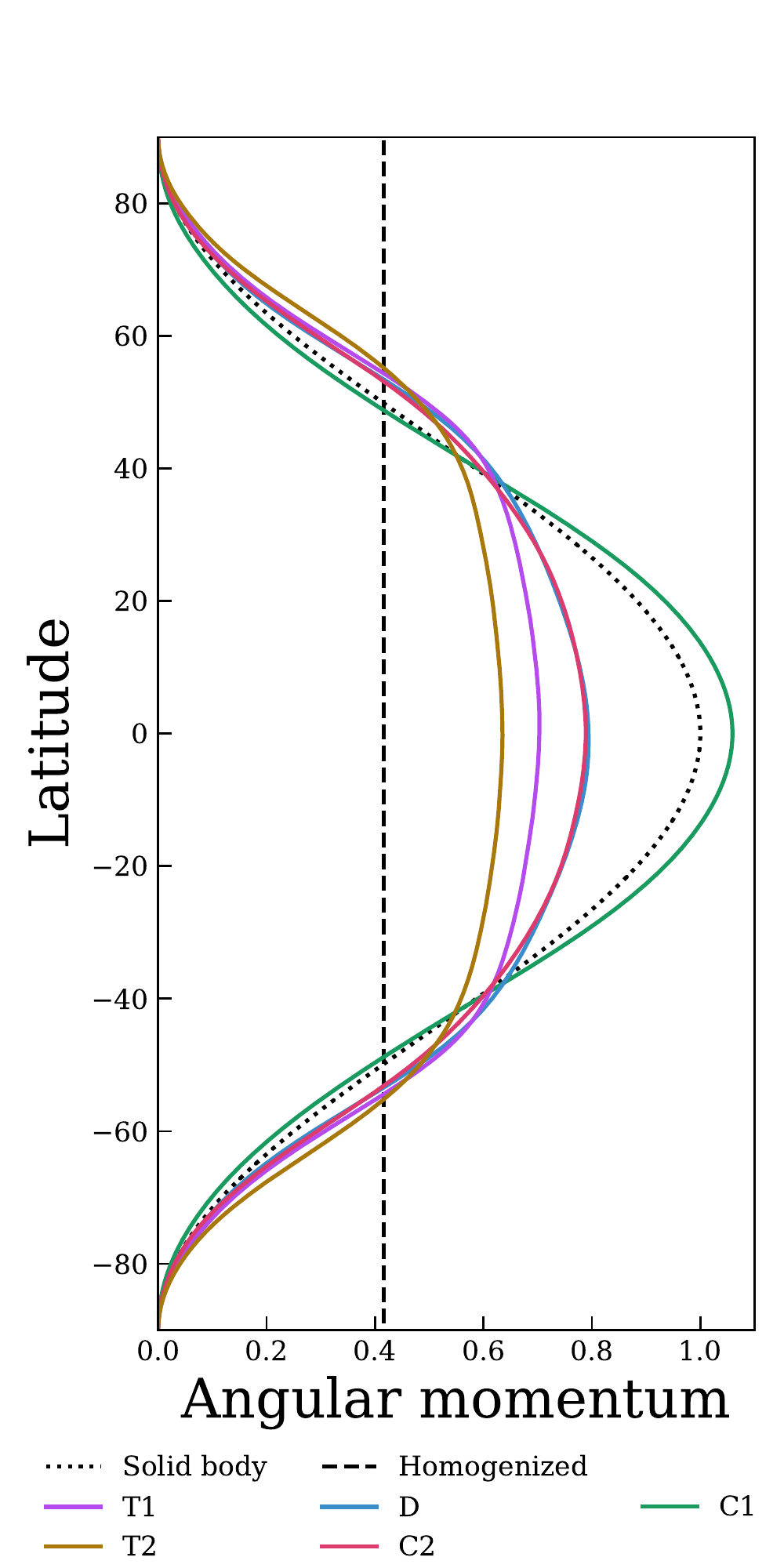}{.375\linewidth}{\large{a)}}
    \fig{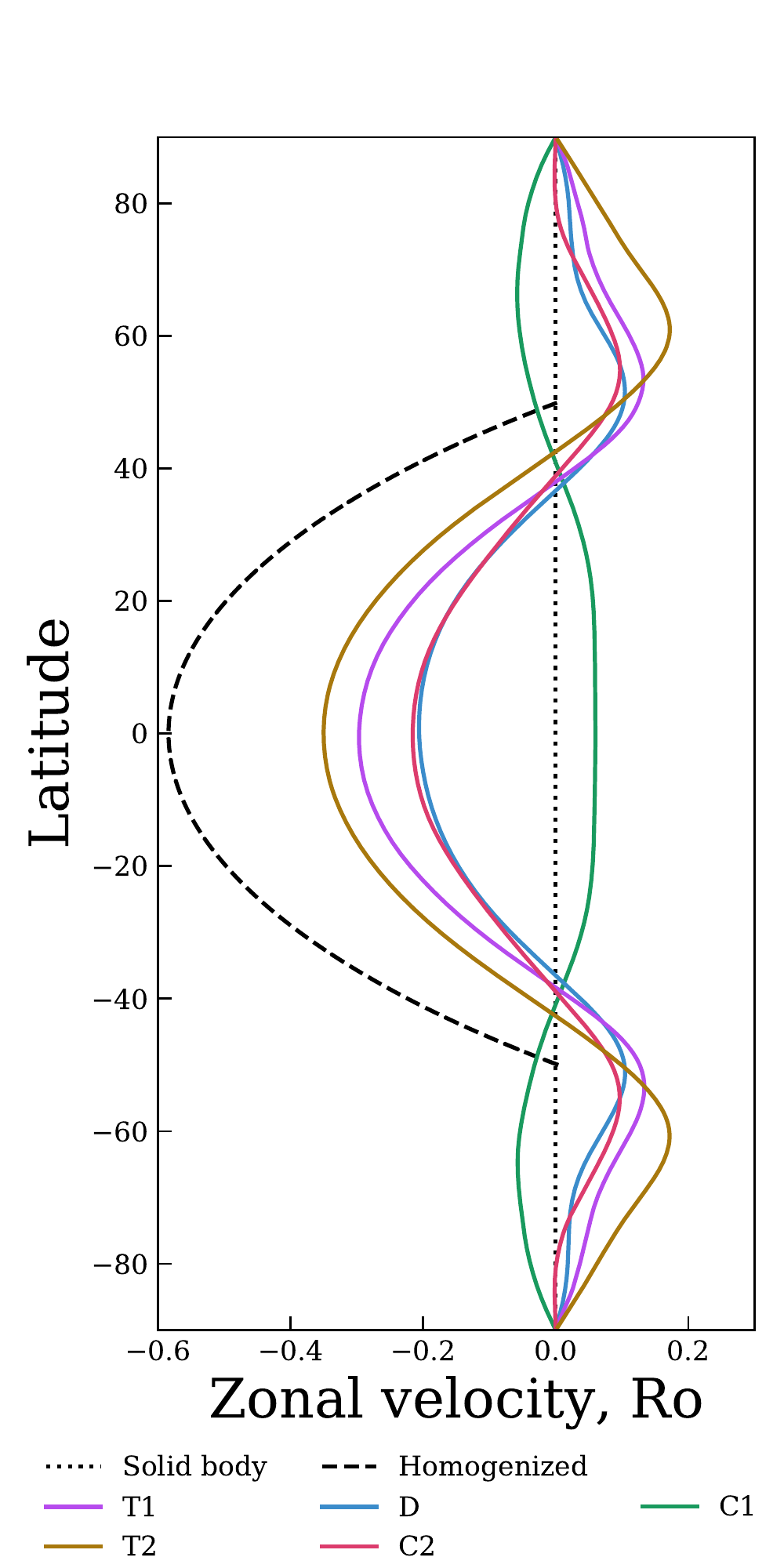}{.375\linewidth}{\large{b)}}
	\caption{a) Axial component of the absolute angular momentum, $h_z$, at the outer boundary as a function of latitude for each case.  The black dashed line represents fully homogenized angular momentum throughout the spherical shell, and the grey dotted curve representing solid body rotation.  b) The zonal flow at the outer boundary for each case, and the predicted zonal flow associated with perfectly mixed angular momentum. }
	\label{fig:latitude_angular_momentum}
\end{figure}

Figure~\ref{fig:latitude_angular_momentum}a plots $h_z$ as a function of latitude at the outer boundary for all models as well as end-member scenarios representing spatially homogeneous angular momentum and solid-body rotation.
For homogeneous angular momentum, the profile is a vertical line.
In regions where the curve is flatter, the gradient is less steep, and therefore we interpret these regions as being more well mixed than where the curve changes more rapidly with latitude \citep{aurnou2007effects,soderlund2013turbulent}.
The thermal cases have quite efficient mixing of angular momentum in their equatorial region, whereas the corresponding compositional cases tend to have less efficient mixing, resulting in a more recurved shape.

Figure~\ref{fig:latitude_angular_momentum}b plots the corresponding zonal flows, including the zonal flow that results from homogenized angular momentum.
In this homogenized case, we expect the formation of not only strong equatorial flows, but also non-physical infinitely fast prograde flows approaching the poles.
In the case of T1 and T2, relatively fast equatorial flows develop, as well as a parabolic shape to the curve around the equator.
In all cases, the flows near the poles are damped out, resulting in local maxima (i.e. prograde bands) at higher latitudes.
However, in the compositional and double-diffusive case, the equatorial flows are significantly diminished compared to the thermal cases.
In C1, rather than seeing well mixed angular momentum, the prograde flow near the equator does not vary strongly with latitude.
Thus, despite identical $Ro_C$ values, a range of angular momentum behaviors are observed, ranging from quite well mixed near the equator in T1 and T2 to poorly mixed and closer to solid body rotation in C1. These mixing behaviors are approached quantitatively in the next section.

\subsection{Degree of mixing \label{sec:mixture}}

A method for quantifying the quality of mixing in a system is the coefficient of variation (CV), which is equal to the standard deviation divided by the mean of a given quantity, which measures the intensity of segregation of the solute \citep{kukukova2009new}.
The mean, $\mu_\psi$, and variance, $\sigma_\psi^2$, of a field $\psi$ are volume weighted.
In the case of a Boussinesq fluid, this choice is equivalent to mass weighted averages.
\begin{align}
    \mu_\psi &= \frac{1}{4 \pi/3 (r_o^3 - r_i^3)} \int_0^{2\pi} \int_0^\pi \int_{r_i}^{r_o} dr d\theta d\phi r^2 \sin \theta \psi(r,\theta,\phi) \label{eq:weighted_mean} \\
    \sigma_\psi^2 &= \frac{1}{4 \pi/3 (r_o^3 - r_i^3)} \int_0^{2\pi} \int_0^\pi \int_{r_i}^{r_o} dr d\theta d\phi r^2 \sin \theta [\psi(r,\theta,\phi) - \mu_\psi]^2 \label{eq:weighted_variance}
\end{align}
\begin{equation}
    CV_\psi = \frac{\sigma_\psi}{\mu_\psi} \label{eq:cov}
\end{equation}
By construction, a CV of zero represents a field that has been fully homogenized.
We are interested in the mixing of temperature, composition, and the axial component of absolute angular momentum.
For each case, time-averaged CV quantities are obtained by averaging CV values calculated for snapshots of the three-dimensional velocity, temperature, and composition fields with the numbers of snapshots used given in Table~\ref{tab:derived_parameters}.
To give context to these results, we also calculate the CV for three special cases: the temperature or composition transported by diffusion alone, the angular momentum of a rotating solid body, and for a fluid that moves as a rotating solid body inside the tangent cylinder but is fully mixed outside the tangent cylinder ($CV_{TC}$).

\begin{figure}
    \fig{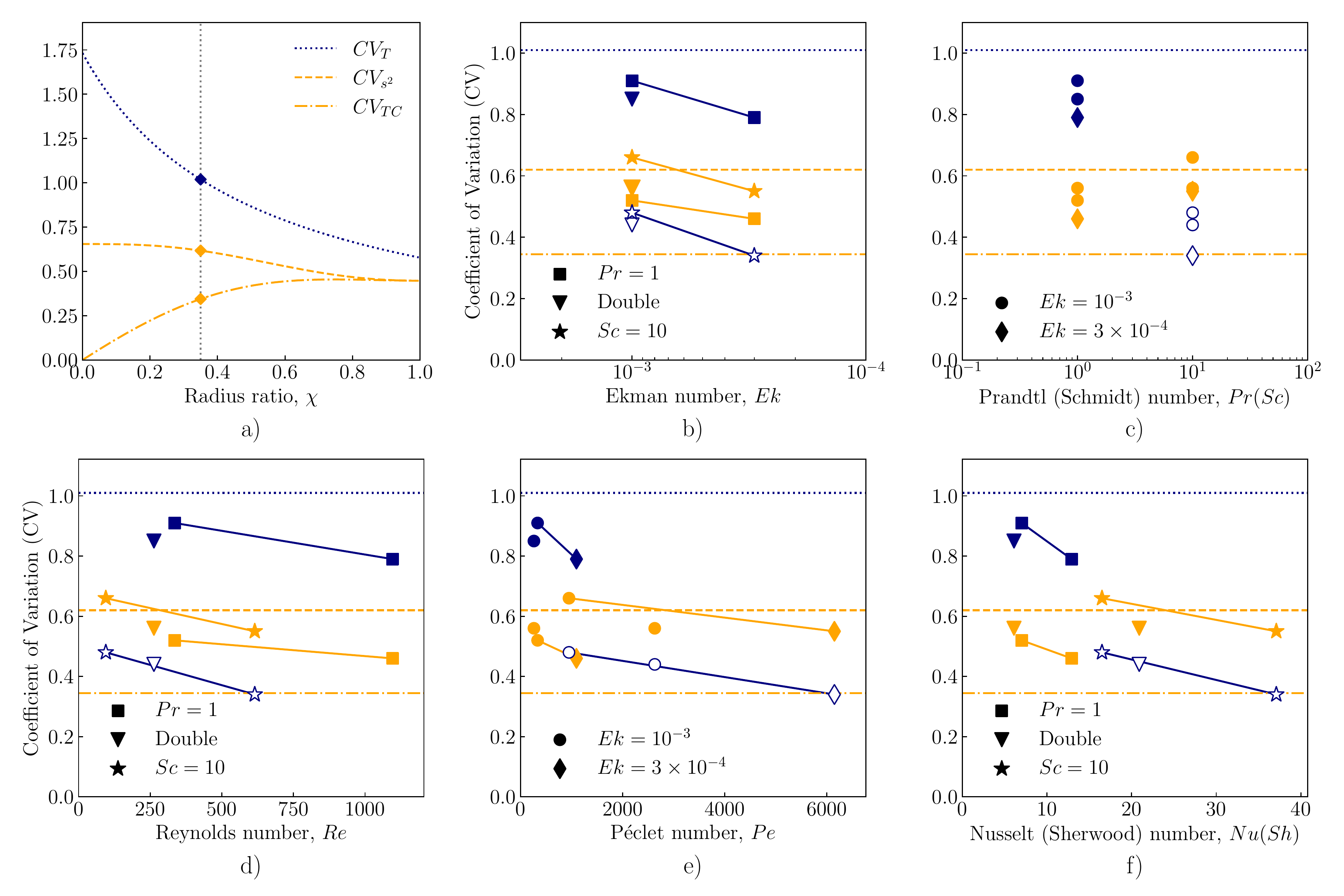}{\linewidth}{}
    \caption{Coefficients of variation vs. a) Ekman number, b) Prandtl (Schmidt) number, d) Reynolds number, e) P\'eclet number, and f) Nusselt (Sherwood) number.
    Lines connect markers from cases with similar convective driving (i.e. T1 to T2 and C1 to C2).
    In a), d) and f), square markers indicate thermal ($Pr = 1$) cases, triangles indicate double diffusion, and starts indicate compositional cases ($Sc = 10$).  In b) and e), circular makers indicate $Ek = 10^{-3}$ cases and diamonds indicate $Ek = 3 \times 10^{-4}$ cases.
    Gold markers indicate absolute angular momentum, filled blue barkers indicate temperature, and empty blue markers composition.
    In b), e) and f), the same $CV_{h_z}$ for the double diffusive case is plotted at both $Pr=1$ and $Sc=10$; however, $CV_T$ is plotted only at $Pr=1$ and $CV_\xi$ is plotted only at $Sc=10$.  Horizontal lines correspond to heat (mass) transport by diffusion (blue dotted), solid body rotation (gold dashed), and tangent cylinder models (gold dashdot).  c) Coefficients of variation as a function of $\chi$, where the vertical grey line marks $\chi = 0.35$.  The diamonds marking the intersection of these curves with the vertical line represent the values of the horizontal lines in the other subfigures.}
    \label{fig:cov}
\end{figure}

In the case of heat transport by diffusion, the system reaches steady state when $\nabla^2 T = 0$ with boundary conditions $T(r_i)=1$ and $T(r_o)=0$ such that the temperature profile will only be a function of normalized radius for a given shell geometry:
\begin{equation}
    T(r) = \frac{\chi}{(1-\chi)^2} \left(\frac{1-(1-\chi) r}{r} \right),
\end{equation}
\noindent where $r_i = \chi/(1-\chi)$ and $r_o=1/(1-\chi)$ (with equivalent arguments for the compositional field). As a result, the mixing properties of these diffusive profiles are:

\begin{equation}
    \mu_T = \mu_\xi = \frac{\chi (2 \chi + 1)}{2 (\chi^2 + \chi + 1)},
\end{equation}
\begin{equation}
    \sigma^2_T = \sigma^2_\xi = \frac{3 \chi^2}{4 (\chi^2 + \chi + 1)^2},
\end{equation}
\begin{equation}
    CV_T = CV_\xi = \frac{\sqrt{3}}{2 \chi + 1}.
\end{equation}
\noindent Figure~\ref{fig:cov}a shows a plot of $CV_T$ for all possible values of $\chi$.
For $\chi=0.35$, the $CV_T$ is equal to 1.02, which establishes an upper bound on cases where fluid motions cause the heat/light elements to be mixed.

Angular momentum of a rotating solid body can be obtained by setting $u_\phi$ equal to zero in Equation \ref{eq:angular_momentum}, leading to
\begin{equation}
    \mu_{s^2} = \frac{2}{5} \,\frac{1 - \chi^5}{1 - \chi^3},
\end{equation}
\begin{equation}
    \sigma^2_{s^2} = \frac{4}{175} \frac{3 (\chi^7 - 1) (\chi^2 + \chi + 1) - 7 \chi^3 (\chi^2 - 1) (\chi + 1)}{(\chi^2 + \chi + 1)(\chi^3 - 1)},
\end{equation}
\begin{equation}
    CV_{s^2} = \frac{1}{\sqrt{7}} \sqrt{\frac{3 (\chi^7 - 1) (\chi^3 - 1) (\chi^2 + \chi + 1) - 7 \chi^3 (\chi^3 - 1) (\chi^2 - 1) (\chi + 1)}{(\chi^2 + \chi + 1) (\chi^5 - 1)^2}}.
\end{equation}
\noindent As shown in Figure~\ref{fig:cov}a, the CV ranges from 0.65 for the full sphere ($\chi=0$) to 0.45 as the spherical shell becomes infinitesimally thin ($\chi \rightarrow 1$).
When $\chi=0.35$, CV is equal to 0.62.
For relatively thick shells ($\chi \lesssim 0.4$), the CV deviates from the full sphere value proportionally to $\chi^3$:
\begin{equation}
    CV_{s^2} = \sqrt{\frac{3}{7}} - \frac{1}{2} \sqrt{\frac{7}{3}} \chi^3 + \mathcal{O} (\chi^4).
\end{equation}
\noindent For thinner shells ($\chi \gtrsim 0.4$), the deviation from the infinitely thin shell value is proportional to $(\chi - 1)^2$:
\begin{equation}
    CV_{s^2} = \sqrt{\frac{1}{5}} + \frac{(1 -\chi)^2}{\sqrt{5}} + \mathcal{O} ((1 - \chi)^3).
\end{equation}

As shown in Figure~\ref{fig:cov}b, decreasing the Ekman number increases the efficiency of both heat (mass) and angular momentum mixing due to the increased intensity of the flows that mix these fields.
This is supported by the general downward trend with increasing Reynolds number seen in Figure~\ref{fig:cov}d.
This trend is also true when controlled for convective driving by comparing T1 to T2 and C1 to C2.

Varying the convective driving has the opposite effect in angular momentum as it does in heat (mass), as shown in Figure~\ref{fig:cov}c.
The reduction of CV from the diffusive end-member value of 1.02 demonstrates the role that convective transport plays in homogenizing the fluids.
Figure~\ref{fig:cov}f shows that while there is not a distinct trend in the angular momentum mixing, increasing the Nusselt (Sherwood) number is associated with more efficient mixing of heat (mass).
The P\'eclet number is the ratio of advective to diffusive transport and given by $Pe_T = Re Pr$ in the case of heat and $Pe_\xi = Re Sc$ in the case of light elements.
As seen in Figure~\ref{fig:cov}e, there is similarly no distinct trend in the angular momentum mixing, but an increase in the Reynolds (P\'eclet) number is associated with a decrease in the CV of heat (mass).
Despite generally weaker flows in the compositional and double diffusive cases compared to the thermal cases, the relative importance of advective heat (mass) transport to diffusion increases with decrease in diffusivity of mass compared to heat.

This can be understood through a feedback process, by which the reduction of diffusivity allows the system to achieve the same degree of heat (mass) mixing with less intense fluid motion, and that a more thoroughly mixed temperature (composition) field causes the influence of buoyancy on the dynamics of the system to become reduced.
We can quantify this one way by considering the total non-dimensional buoyant power in each case \citep{yadav2013consistent}:
\begin{equation}
    P =  \int dV u_r \left( \frac{Ra}{Pr} T + \frac{Ra_\xi}{Sc} \xi \right) g(r).
\end{equation}
Comparing T2 and C2, we find that T2 has a buoyant power of $1.9 \times 10^9$ compared to $5.8 \times 10^8$ in C2 (only 30\% of T2).
A similar decrease occurs from T1 to D to C1, which have buoyant powers of $8.9 \times 10^7$, $5.2 \times 10^7$ and $2.2 \times 10^7$, respectively.
This means that as the convective driving is varied to less diffusive modes of convection, the buoyant force is contributing less to the energetics of the system, and therefore the strength of the flow is weakened.
Despite this, efficient mixing of mass is maintained.

\subsection{Force balance}

The dominant force balance determines the evolution of the dynamics in the system.
The force terms of Equation (\ref{eq:navier_stokes}), $\mathbf{f}$, can be expressed in terms of components in spherical coordinates,
\begin{equation}
    \mathbf{f} = f_r \hat{\textbf{r}} + f_\theta \hat{\boldsymbol{\theta}} + f_\phi \hat{\boldsymbol{\phi}}.
\end{equation}
The root-mean-square (RMS) value as a function of spherical harmonic degree $n$ may be time- and volume- averaged as
\begin{equation}
    (f^n_{rms})^2 = \sum_{m=0}^n \frac{1}{V \Delta t} \int dt \int dV (f^2_r + f^2_\theta + f^2_\phi) Y^m_n(\theta,\phi),
\end{equation}
\noindent where $\Delta t$ is the duration of the time-averaging.
The global value may then be obtained by summing over all spherical harmonic degrees:

\begin{equation}
    (f_{rms})^2 = \sum_{n=0}^{n_{max}} (f^n_{rms})^2.
\end{equation}

\begin{figure}
    \fig{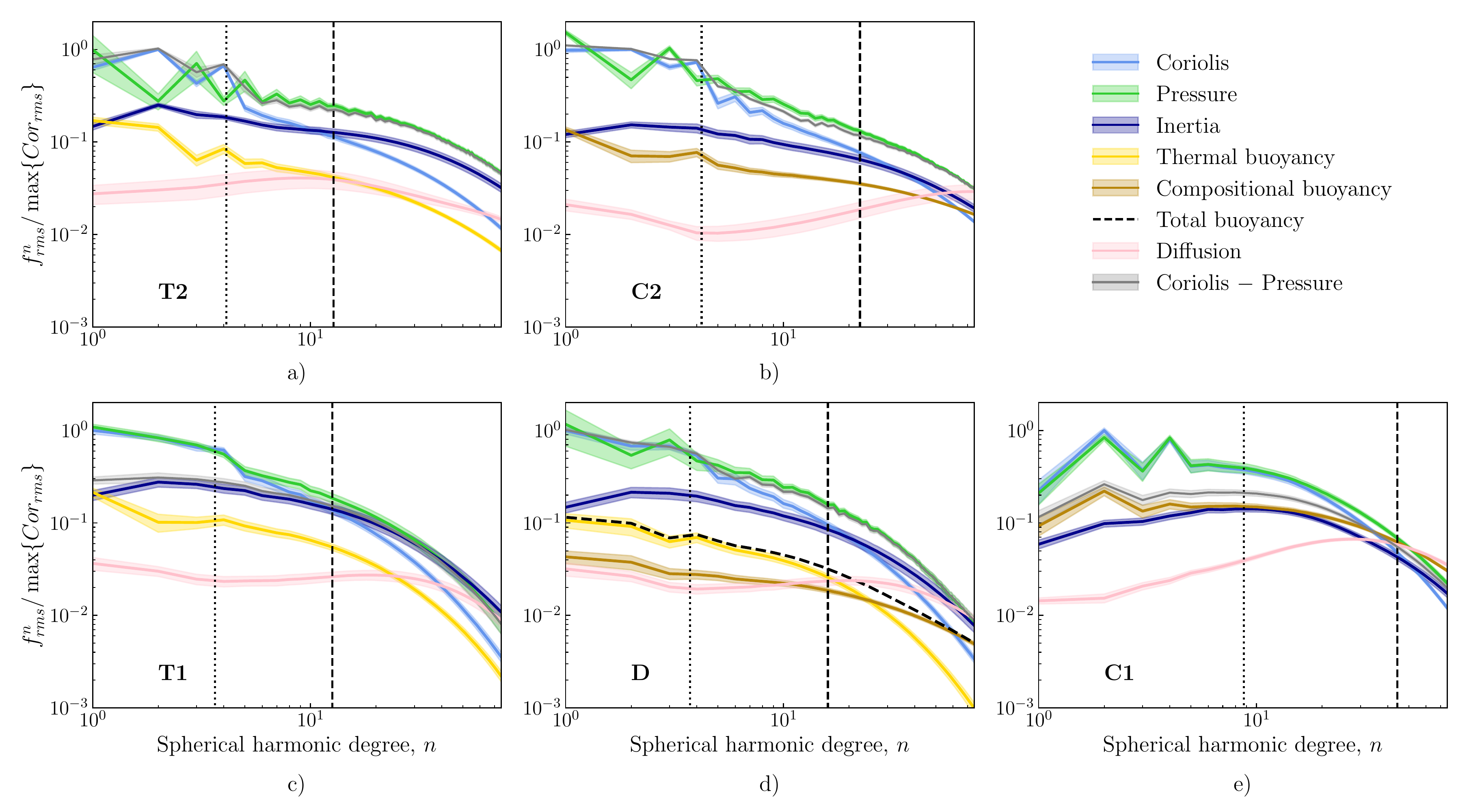}{\linewidth}{}
 	\caption{Root-mean-square terms of Equation \ref{eq:navier_stokes} as a function of spherical harmonic degree, $n$, up to $n = 75$.  The left vertical axis shows the value normalized by maximum value of the Coriolis force.  The vertical black dotted line represents the mean length scale in Equation \ref{eq:mean_degree}, and the vertical grey dashed line represents an empirical length scale proportional to the Rossby number given in Equation \ref{eq:crossover}, where the Coriolis force should cross with inertia.}
	\label{fig:force_balance}
\end{figure}

Figure~\ref{fig:force_balance} plots of the time-averaged magnitude of forces in the system throughout the full shell.
At small $n$, the most dominant forces are the pressure gradient and Coriolis in all cases; however, different forces become prominent at smaller length scales.
In T1, the ageostrophic Coriolis term exceeds the magnitude of the Coriolis force at $n \gtrsim 11$, which indicates that a substantial amount of the Coriolis force is anti-parallel to the pressure gradient.
As $n \rightarrow n_{max}$, the force balance becomes dominated by viscosity.
In C1, there is still a leading order geostrophic balance at length scales larger than the characteristic wavenumber, with pressure and Coriolis force peaks at $n=2$ and 4 that substantially exceed the $n=1$ values in contrast to the other cases.
As $n \rightarrow n_{max}$, while the viscosity again becomes the greatest force in the force balance, the compositional buoyancy remains relatively strong because smaller scale features of the composition field persist.
Case D maintains properties similar to both T1 and C1, passing through configurations like both cases with increasing $n$.
The strength of the compositional buoyancy, though it begins at small $n$ with a low value, persists to higher spherical harmonic degrees, crossing over the thermal buoyancy at $n \approx 24 $.
Viscosity crosses inertia at $n \approx 62$.
At this point, the dominant force balance resembles T1 more, but compositional buoyancy crosses inertia at $n > 100$.

In T2 and C2, there are peaks in pressure at $n=1$, 3, and 4 and in the Coriolis force at $n =2$ and 4.
Above $n = 5$, pressure continues to dominate the force balance of T2.
Inertia crosses over the Coriolis force at $n=10$.
The viscosity only eventually becomes comparable to the pressure and inertial forces as $n \rightarrow n_{max}$.
In C2, inertia crosses over the Coriolis force at $n \approx 35$, however as $n$ increases, the Coriolis force eventually becomes the least significant force.
At $n \approx 80$, viscosity crosses over the pressure, where it becomes the dominant force in the system.

In each case, inertia crosses over the Coriolis force as $n$ is increased.
We can attempt to find where this crossing occurs by assuming that the Coriolis term is similar in strength to inertia at the length scale $\ell_{CI}$:

\begin{equation}
    2 \boldsymbol{\Omega} \times \mathbf{u} \sim \mathbf{u} \cdot \nabla \mathbf{u},
\end{equation}
which can be recast as
\begin{equation}
    \ell_{CI}/D \sim Ro = Ek Re (1 - \chi). \label{eq:crossover}
\end{equation}
\noindent Because our models tend to have $Ro \sim 0.1$, this crossing is expected to occur at moderate values of $n$ consistent with Figure~\ref{fig:force_balance}.
The crossover length can be obtained from the spectra by comparing the magnitude of the Coriolis force and inertia.
Because we expect a linear relationship between the Rossby number and this length scale, we are able to obtain an empirical linear fit of the form $\ell_{CI} = a Ro$ between these quantities using the least-squares method.
The fit we obtain is
\begin{equation}
    \ell_{CI} / D \sim 0.75 Ro.
\end{equation}
These values are plotted in Figure~\ref{fig:force_balance} as the vertical dashed lines.
This prediction is more successful in the thermal cases, where these lines pass almost exactly through the crossover, than in the compositional and double diffusive cases.
However, in the non-thermal cases, the crossover occur at degrees where other forces have become significant.
This is especially true in C1, where all the forces are approximately equal to one another.

The magnitude of the force terms in Equation~\ref{eq:navier_stokes} are dependent on the choice of characteristic time scale, which depends on the choice of viscosity.
However, if instead we choose rotational time as the characteristic time, we are able to compare the magnitudes of the forces between each case, assuming that the rate of rotation is the same between them.
This is equivalent to considering the dimensional forces to be normalized by $D \Omega^2$
Values for the integrated forces, as the viscosity and convective driving is varied are tabulated in Table~\ref{tab:force_balance}.
First, the magnitudes of buoyancy are less effected by decreasing $Ek$ than by varying the convective driving.
Comparing the compositional cases to the thermal cases, the compositional buoyancy falls off less rapidly with increasing $n$ than thermal buoyancy, but thermal buoyancy has a higher peak value than compositional buoyancy.
This has the effect of reducing the integrated buoyant force in the compositional cases.
In T2, the integrated buoyant force is 0.14, compared to 0.07 in C2.
In T1, the integrated buoyant force is 0.15, compared to 0.09 in D and C1.
When this is taken with the discussion of the buoyant power in Section \ref{sec:mixture}, as the heat or mass become more thoroughly mixed, the influence of the buoyancy is reduced.
As the system undergoes weaker forcing, the velocity field decreases in intensity.

\begin{figure}
    \fig{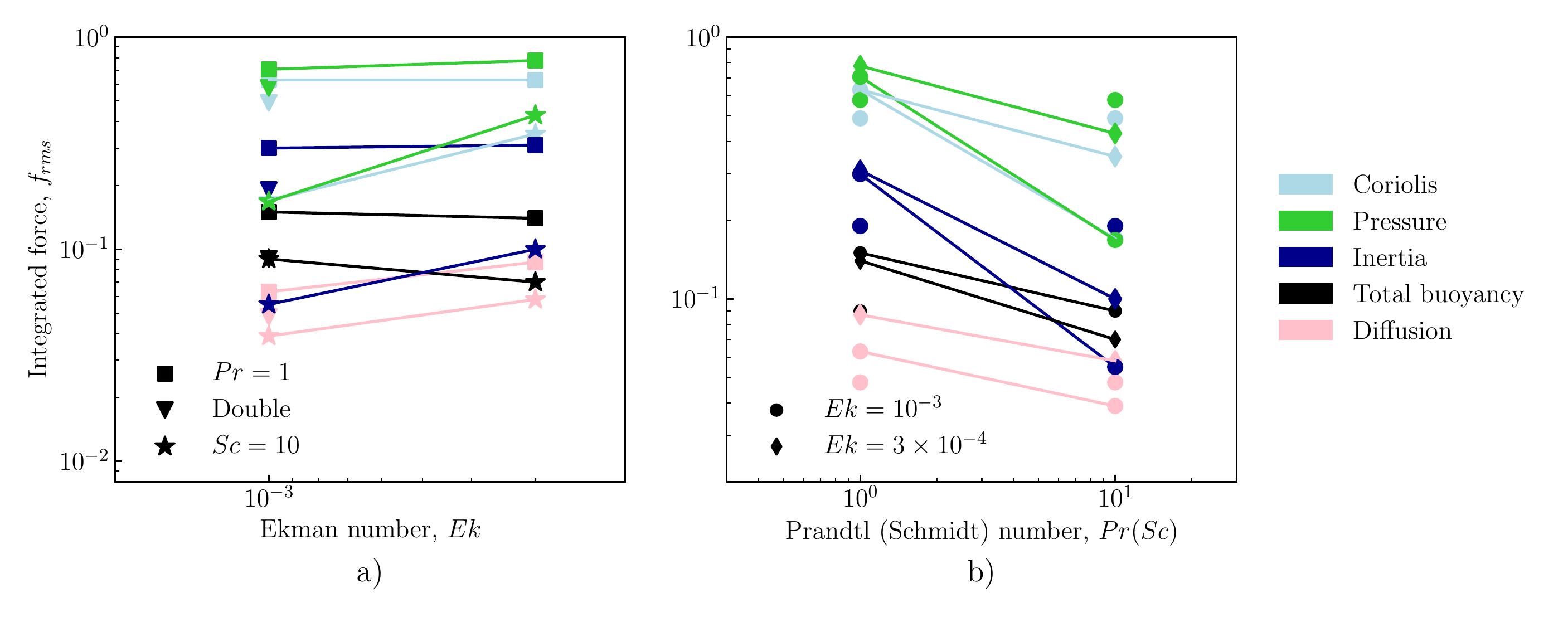}{\linewidth}{}
    \caption{Integrated RMS force terms vs. a) Ekman number and b) Prandtl (Schmidt) number.  Because the shell thickness and rate of rotation of the systems are constant, the forces are normalized by $D \Omega^2$.}
    \label{fig:integrated_force}
\end{figure}

Figure~\ref{fig:integrated_force} contains plots of the globally integrated forces against the input parameters.
The convective Rossby number is intended to be a proxy for the ratio between buoyancy and the Coriolis force, with $Ro_C \sim 1$ often delineating the transition between eastward and westward equatorial flows \citep{aurnou2007effects,gastine2013solar}.
As the Ekman number is decreased, the total buoyancy is roughly consistent to slightly decreased for a given mode of convection.
However, for constant Ekman number, total buoyancy decreases moving from thermal to compositional convection.
While there is no change between T1 and T2, but a notable decrease in the total Coriolis force between C1 and C2.
Furthermore, in cases T1 and T2 and the thermal component of D, buoyancy is never greater than either the Coriolis force nor inertia at any spherical harmonic degree.
In cases C1, C2, and D, the compositional buoyancy force becomes greater than  Coriolis at large values of $n$.
The ratios of the integrated buoyancy to Coriolis force are tabulated in Table~\ref{tab:force_balance}.
In the case of C1, the volume integrated buoyancy compared to Coriolis is the strongest among the different cases with a force ratio of $\sim 0.5$ compared to $\lesssim 0.2$ for the other cases.
Comparing the properties of each system in Figure~\ref{fig:equatorial_slice} shows that the dynamics of T1, T2, D, and C2 tend to be dominated by lower order terms, while the relative importance of higher order terms is reflected in the spectrum for C1, and the peaks that form in the Coriolis, pressure and buoyancy terms at $n=2$ and $n=4$.

\begin{deluxetable}{cccccccccc}
    \caption{Total integrated momentum diffusion, inertial, thermal buoyancy, compositional buoyancy, total buoyancy, pressure gradient, Coriolis, and ageostrophic Coriolis forces, in terms of $D \Omega^2$, and the ratio of buoyancy to Coriolis forces for each case. \label{tab:force_balance}}
    \tablehead{
        \colhead{} & \colhead{} & \colhead{} & \colhead{Thermal} & \colhead{Compositional} & \colhead{Total} & & \colhead{} & \colhead{} & \colhead{} \\
        \colhead{Case} & \colhead{Diffusion} & \colhead{Inertia} & \colhead{Buoyancy} & \colhead{Buoyancy} & \colhead{Buoyancy} & \colhead{Pressure} & \colhead{Coriolis} & \colhead{Coriolis $-$ Pressure} & \colhead{Buoyancy/Coriolis}
    }
    \startdata
    T1 & 0.06 & 0.30 & 0.15    & \nodata & 0.15 & 0.71  & 0.63 & 0.34 & 0.24 \\
    D  & 0.05 & 0.19 & 0.08    & 0.05    & 0.09 & 0.58 & 0.49 & 0.52 & 0.19 \\
    C1 & 0.04 & 0.05 & \nodata & 0.09    & 0.09 & 0.17 & 0.17 & 0.10 & 0.52 \\
    T2 & 0.09 & 0.31 & 0.14    & \nodata & 0.14 & 0.78 & 0.63 & 0.71 & 0.22 \\
    C2 & 0.06 & 0.10 & \nodata & 0.07    & 0.07 & 0.43 & 0.35 & 0.37 & 0.19 \\
    \enddata
\end{deluxetable}

\subsection{Gravitational moments \label{sec:gravitational_moments}} 

Processes occurring within planetary interiors are not directly observable and, therefore, their properties must be inferred from observable properties such as their gravitational fields.
Anomalies in temperature and composition give rise to buoyancy in the system create anomalies in the density.
These anomalies are not necessarily radially symmetric, like the hydrostatic background, meaning that the gravitational field that results from them may likewise not have a radial symmetry.

The gravitational field surrounding a planet is irrotational and can therefore be represented in terms of a scalar potential and multipole expansion.
The strength of each term is determined by the distribution of mass inside the planet and quantified using the gravitational moment, $J_n$,
\begin{equation}
    M r_o^n J_n = \int_0^{2\pi} \int_0^{\pi} \int_0^{r_o} \rho(r, \theta, \phi) r^{n+2} \sin \theta P_n(\cos \theta) dr d\theta d\phi,
\end{equation}
\noindent where $P_n$ is the Legendre polynomial of degree $n$.
When $m=0$, the moments are called the zonal gravitational moments.
These moments can be further divided into two components, arising from the flattening that is associated with solid body rotation and the dynamics of the system, $\Delta J_n$.
Solid body rotation causes flattening that is symmetric with respect to the equatorial plane and, therefore, the gravitational moments associated with odd values of $n$ can only result from the dynamics of the system. In contrast, even moments reflect both planetary structure and dynamics.
Here, we calculate the density anomaly associated with dynamics in our models, noting there are two effects that are not represented: the flattening associated with the rotation of the system and the increase in background density with depth.

The first two even zonal gravitational moments, $J_2$ and $J_4$, were measured in Uranus and Neptune by \textit{Voyager 2} \citep{stone1986voyager,stone1989voyager}.
Gravitational science has remained one of the primary of objectives for more recent scientific missions to the giant planets, with higher order gravitational moments being measured to $J_{10}$ in Jupiter by \textit{Juno} and to $J_{12}$ in Saturn by \textit{Cassini-Huygens} \citep{debras2019new,galanti2019saturn}.
Mission concepts to the ice giants typically seek to constrain the low degree gravitational moments up to at least $J_6$ \citep[e.g.,][]{hofstadter2017ice,jarmak2020quest,rymer2020neptune}.

Equation \ref{eq:boussinesq_approximation} relates the density anomaly to the temperature and composition fields, from which we can derive an 
equation for the gravitational moment resulting from the dynamics of the fluid:
\begin{equation}
    \Delta J_n = \frac{3}{4 \pi} \frac{R^3 \Omega^2}{G M} \frac{(1 - \chi)^{n+1}}{1 + \left( \frac{\chi}{1-\chi} \right)^3 \frac{\bar \rho}{\rho_c}} \int dV Y^m_n(\theta,\phi) \left[ f_T T' (r,\theta,\phi) + f_\xi \xi' (r,\theta,\phi) \right],
\end{equation}
\noindent where $\rho_c$ is density of the core region, and $f_T = Ek^2 Ra/Pr$ and $f_\xi = Ek^2 Ra_\xi/Sc$ are the fractional contributions of thermal and compositional to buoyancy, respectively.

In our calculations, we assume that $\rho_c \approx 10 \bar \rho$ as an order-of-magnitude approximation, which is similar to the ratio of densities in other interior models \citep[e.g.,][]{hubbard1991interior,nettelmann2013new}.
The quantity $R^3 \Omega/G M$ is approximately 0.03 in ice giant planets.
This estimate is obtained by considering reasonable choices of constant density for the fluid and core region to approximate the interior of Uranus, assuming that a distinct transition between these layers exists.
For example, if the density of the fluid is approximately equal to 1000 kg m$^{-3}$, then the density of the core would be approximately 7300 kg m$^{-3}$ so that the mean density of the entire system matches that of Uranus.
Assuming that the density of the fluid is approximately 500 kg m$^{-3}$, the core density would be 18000 kg m$^{-3}$ (36 times larger) so that the mean density matches Uranus.
Decreasing the choice of density in the fluid would require a likely un-physically large density in the core while a fully uniform interior would result in a ratio of one, with a choice intermediate between these extremes more likely to approximate the physical reality of the planetary interiors.

\begin{figure}
    \fig{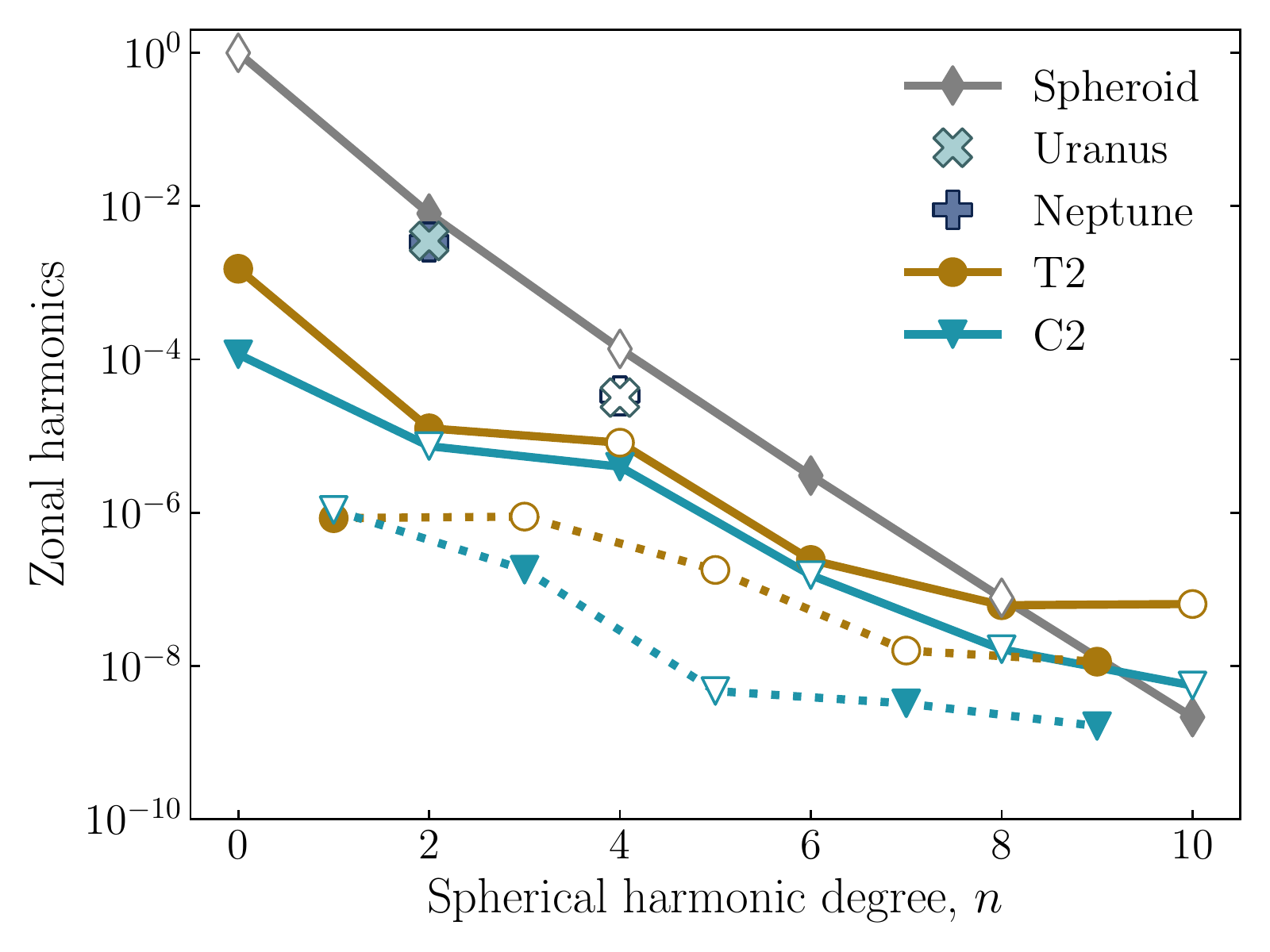}{.49\linewidth}{\large{a)}}
    \fig{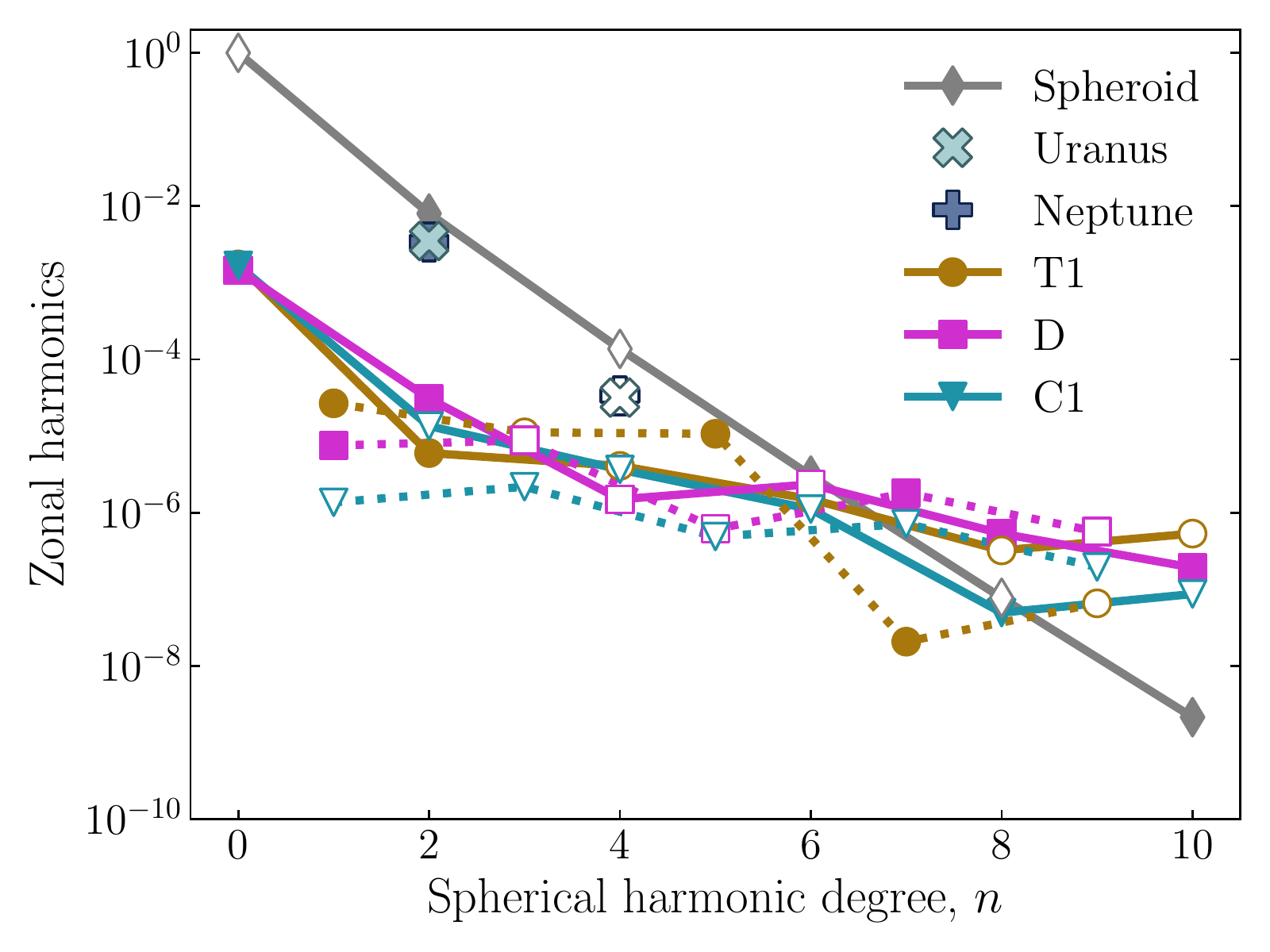}{.49\linewidth}{\large{b)}}
    \caption{Zonal gravitational moments for a) $Ek = 3 \times 10^{-4}$ and b) $Ek = 10^{-3}$ cases.  Filled markers indicate a positive value and empty markers indicate negative values.  In the calculated zonal gravitational moments, solid lines connect the even series, and dotted lines connect the odd series.  The grey dotted line represents the even gravitational moments that result from a rotating, uniform spheroid with flattening that approximates the ice giants \citep{hofmeister2018verified}.  Measured gravitational moments, $J_2$ and $J_4$ are also included for Uranus and Neptune, though the values are very similar and overlap one another \citep{jacobson2009orbits,jacobson2014orbits,williams2015nasa}. \label{fig:gravitational_moments}}
\end{figure}

Figure~\ref{fig:gravitational_moments} plots the time-averaged gravity spectrum up to spherical harmonic degree 10.
Though we have ignored flattening due to rotation, gravitational moments associated with a uniform spheroid with flattening similar to the ice giants has been included to give context to the dynamical gravitational moments calculated from the simulation results.
Calculations for $J_{2n}$ are provided by \cite{hofmeister2018verified}:
\begin{equation*}
    J_{2n} = \frac{-3 (-1)^n e^{2n}}{(2n + 1)(2n+3)},
\end{equation*}
\noindent where $e$ is the ellipticity of the spheroid.
Similar to the coefficients of variation, gravitational moments are calculated from snapshots of the temperature and composition fields, then indexed temporally to create a time series.
For each radial level, the spectral representation of temperature and composition are determined, which are then integrated to determine the values of $\Delta J_n$ by reconstructing the spherical harmonic representation of the field and relying on the orthogonality of these functions \citep{holmes2002unified,wieczorek2018shtools}.
From this time series, the statistical properties of the gravitational moments can be calculated.
At low values of $n$, this solid body rotation produces a stronger gravitational signature than the dynamics in the system, but also falls off  more rapidly.
As a result, the dynamics produce a stronger signal at $n \gtrsim 6$ when $Ek=10^{-3}$ (Figure~\ref{fig:gravitational_moments}b) and $n \gtrsim 8$ when $Ek=3\times10^{-4}$ (Figure~\ref{fig:gravitational_moments}a). 
Some insights can be gained into these results by considering the features of the temperature and composition fields in Figure~\ref{fig:temperature_slices}.
There is a reduction in the mean odd moments as the Ekman number is decreased.
Comparing T1 to T2 and C1 to C2, the equatorial bulge in temperature and composition are larger in T1 and C1.
That being said, these moments are composites of many snapshots in time, and we see that this bulge is, in most of the cases, associated with the non-symmetric features from the right column of Figure~\ref{fig:equatorial_slice}.
Comparing T1 (Figure~\ref{fig:equatorial_slice}i) to T2 (Figure~\ref{fig:equatorial_slice}c), the temperature anomalies extend higher in the spherical shell, where the gravitational moments will be more sensitive.
Furthermore, due to the three-dimensional nature of these flows, equatorial asymmetries will develop that are probed by the gravitational moments.

In the case of $\Delta J_2$ and $\Delta J_4$, the variance of the temperature and composition fields and the resulting uncertainty of the gravitational moments in our models is only moderate, but the standard deviation of the gravitational moments increases with $n$.
The time series of the gravitational moments are stationary, however, meaning that the values presented in Figure~\ref{fig:gravitational_moments} do not change over the course of the simulation.
In case T1 and T2, only values of $\Delta J_2$ and $\Delta J_4$ were statistically non-zero (determined by a students-t test), but in the case of D only, $\Delta J_2$ was non-zero.
In cases C1 and C2, the harmonics $\Delta J_2$, $\Delta J_4$, $\Delta J_6$, and $\Delta J_{10}$ were statistically non-zero.
For odd zonal gravitational moments, values of the calculated harmonics tend to alternate between positive and negative values, such that the time-average is not statistically different from zero because the non-equatorially symmetric features of the temperature and composition tend to alternate between higher concentrations in the northern and southern hemisphere over time.

\section{Discussion and Conclusions\label{sec:discussion}}

We performed a suite of rotating convection models driven by thermal and/or compositional buoyancy sources at different Ekman numbers for a fixed convective Rossby number to probe hypothesized processes in ice giant planets.
Most of the cases develop similar flow behaviors, including the formation of large overturning circulation cells and three zonal jets, but there is more diversity in the thermal and compositional behaviors.
Equatorial peaks in thermal and mass fluxes appear in T1, D, and C2, with diminished equatorial flux in the case of T2 and C1 (see Figures \ref{fig:zonal_flow} and \ref{fig:flux_latitude}).
Case C1 ($Ek=10^{-3}$, $Sc=10$) exhibits significantly different behavior.
While this case has three zonal jets, their directions are reversed with prograde equatorial flows and retrograde bands at high latitudes; it similarly lacks an equatorial peak in mass flux that is likely caused by the disruption of circulation cells in the region near the equator.
Reducing viscosity ($C1 \rightarrow C2$ and $T1 \rightarrow T2$) tends to improve the efficiency of absolute angular momentum mixing, which produces and/or amplifies a retrograde equatorial jet (Figure~\ref{fig:latitude_angular_momentum}).
Reducing the diffusivity of the thermodynamic field ($T1 \rightarrow C1$ and $T2 \rightarrow C2$) tended to produce zonal flows that adhered more closely to solid-body rotation, i.e. with weaker flow speeds (Figure~\ref{fig:latitude_angular_momentum}).

\subsection{Zonal flows and mixing}

Physical properties of the fluid in the interior of planets, such as viscosity and thermodynamic diffusivities, are more extreme than those used in our models, but we can try to predict the effect of scaling these parameters to planetary values based on how well the angular momentum and thermodynamics might be mixed as discussed in Section \ref{sec:mixture}.
The main determinant of how well mixed the thermodynamics became in our models was the choice of diffusivity (Figure~\ref{fig:cov}c).
Planetary values of thermodynamic diffusivities are much smaller than those assumed in our models, likely even relative to the viscosity, and therefore the ice giant interiors may be more thoroughly mixed in terms of the temperature and composition.
Reducing the viscosity causes the flow to become more turbulent, which allows the fluid to become more well mixed (Figure~\ref{fig:cov}b).
Continuing to reduce the viscosity to planetary values while maintaining convection characterized by 3D turbulence should result in increasingly well mixed systems, eventually exhibiting nearly uniform thermodynamic fields in the bulk and zonal winds that approach what is predicted by fully homogenized absolute angular momentum.
In the case of $\chi = 0.35$, full homogenization of angular momentum results in extreme equatorial jet speeds of 1500 m s$^{-1}$, and realistic speeds are not predicted for any spherical shell geometry if full homogenization is assumed.
Restriction of angular momentum homogenization to occur only outside of the tangent cylinder still results in westward zonal speeds of 1300 m s$^{-1}$.
The equatorial wind speeds of Uranus and Neptune can be be reproduced by fully-homogenized angular momentum mixing for $\chi = 0.99$ and $\chi=0.87$, respectively, with thicker domains possible if only partial mixing were to occur or if full mixing of some different initial value of total angular momentum are assumed. 

In contrast, decreasing the thermodynamic diffusivity has an antagonistic effect (Figure~\ref{fig:cov}c) on the absolute angular momentum mixing, which reduces the intensity of zonal winds throughout the system.
Because of the complex interactions between these effects, it is conceivable that with a sufficiently low thermal or mass diffusivity, that the system could be nearly perfectly homogenized in heat or mass, but still maintain a relatively poor state of angular momentum mixture, preventing the formation of winds as intense as those predicted by perfect angular momentum mixing.

These velocity fields emerge from dynamics that are, for the most part, intended to represent convection that is only weakly influenced by the effects of rotation.
Global geostrophic balance is present in neither T1 nor T2, nor is it present in D or C2 (Table~\ref{tab:force_balance}), with this type of balance present only at the largest length scales of the system.
However, this is not the case in C1, where the integrated  pressure gradient and Coriolis force are roughly balanced.
Furthermore, the convective Rossby number failed to predict the direction of the zonal jet and the ratio of integrated total buoyancy to the Coriolis in case C1.

The density anomalies used to calculate the gravitational moments give rise to buoyancy that drives the flows within the system, meaning that we would expect similarities in the flows to be reflected in similarities between the gravitational moments of the system.
Decreasing the Ekman number in the thermal cases does not result in significantly different zonal flows as shown in Figure~\ref{fig:zonal_flow}, and despite the fact that there is a reversal of direction, the flows in the compositional cases have similar morphology.
The magnitude of $J_2$ and $J_4$ tend to be similar as Ekman number is decreased as well as when the convective driving is varied.
However, there are sign changes evident when comparing $J_2$ between the thermal and compositional cases and for $J_4$ in the case of T2 and C2.
These gravitational moments are those that arise only from the dynamics of the fluid contained within the shell.
If the effects of a spheroidal geometry can be ignored, then these flows would modify the gravitational signature of an oblate spheroid in a way such that C2 is more representative of the ice giants than T2.
It would be useful to study the effects of decreasing the diffusivity of heat and light elements further to learn if this trend continues.

\subsection{Light element and heat fluxes}

The patterns of light element fluxes in our compositional and double diffusive cases may also be compared against the latitudinal variations in composition that have been observed in the ice giants.
In particular, methane and hydrogen sulfide
appear to have greater relative abundances in polar regions, although these observations are based on infrared observations of the ice giants that probe the top few bars of the atmosphere \citep[e.g.,][]{moses2020atmospheric}.
Analogously to how sea water becomes brackish from the influx of freshwater near the mouth of a river, the measured deficiency of methane relative to other atmospheric gases may result from an influx of molecular hydrogen.
We assume that there are no additional chemical reactions occurring in the deep atmosphere as the light elements are transported from the interior.

While the flux of light elements is large in the polar regions in the case of C1, this feature appears to be caused by the disruption of equatorial circulation cells that reappear as the viscosity is decreased.
In the case of C2, local maxima in light element flux appear in the polar regions, but are still accompanied by strong equatorial upwellings that transport mass upward from the interior.
The injection of a greater amount of light elements could affect the composition of the atmosphere, and while C1 may be consistent with the observed methane depletion in the polar regions of the ice giants, the equatorial peak in C2 implies that the composition of the atmosphere near the equator should be similarly affected.
In the double diffusive case, the equatorial maxima also disappear, and the mass flux becomes dominated by an equatorial peak.
For this reason, none of our cases are able to account for both the dynamical properties and the atmospheric compositions in a self consistent manner.
Furthermore, based on trends with the P\'eclet number seen in our simulations, we may expect the mass flux to become more uniform with latitude when approaching planetary values of both the Reynolds number (which is very large, assuming a turbulent interior) and Schmidt number.
Given this dissimilarity and the simplicity of our chemistry assumptions, additional observations and determination of high pressure-temperature material properties will be necessary to better understand the chemistry of the deep atmosphere.

While the internal heat flux of Neptune has local maxima at the equator and at the poles, the heat flux of Uranus appears to be less variable with latitude \citep{pearl1990albedo,pearl1991albedo}.
The simulated heat flux patterns tend to have significant latitudinal variability with distinctive peaks at the poles and a secondary peak at the equator (T2), a distinctive peak and the equator and secondary peaks at high latitudes (T1 and D). 
Though Neptune's equatorial peak is larger than either polar peak, none of these cases are particularly reminiscent of the observed flux pattern. 
The change in simulated heat flux pattern as the Ekman number is reduced, however, indicates that further work is needed to determine how these results scale toward more realistic parameters.
Similarly, none of the cases produce results that appear similar to the heat flux of Uranus.
The mass flux of C2 is the least variable of all the cases and bears the greatest similarity to the heat flux of Neptune, suggesting that similar parameters (e.g., $Pr = 10$) or perhaps a double diffusive case with a small relative contribution of thermal driving (e.g. $f_T \leq 25\%$) may produce a heat flux with less latitudinal variability.

Tropospheric processes are not considered in our models, but the time scales involved in atmospheric transport are likely to be much faster than in interior processes.
In the context of shallow atmospheric dynamics, mass and heat flux from the deep interior may be potentially significant forcing terms and this interior-atmosphere coupling is another exciting direction for future inquiry.

\subsection{Density stratification}

We assume that the fluid in our models is incompressible, which neglects any background density stratification, and that the outer boundary represents the 1 bar surface level in order to better facilitate comparisons with observations.
Note that an alternative interpretation is that the outer boundary corresponds to the top of the ionic ocean, in which case the link to observations is obscured. 

Interior models suggest that the density contrast across the ionic ocean is expected to be relatively moderate with much more rapid density changes closer to the surface \citep[e.g.,][]{helled2020uranus}.
The density contrast between the inner and outer boundaries can be quantified by the number of density scale heights, $N = \log (\rho_i/\rho_o)$.
Model U1 of \cite{nettelmann2013new}, for example, has a rapid decrease with increasing radius above $r = 0.9 r_o$.
If we approximate the outer density as $\rho_o \approx 100$ kg m$^{-3}$ at $r \approx 0.95 r_o$ and the inner density as $\rho_i \approx 4000$  kg m$^{-3}$ at $r \approx 0.3 r_o$, then about 3.6 scale heights exist across the shell.
However, if the convective region of interest is confined to the ionic ocean, then the increase in density across this layer is not predicted to be very large, perhaps only representing a single scale height \citep{nettelmann2013new}.

The effect of anelasticity on the dynamics of the system is reflected primarily in the overturning circulations, which become concentrated in a layer near the surface; while the speed of zonal flows has been found to increase in the non-rotating regime of convection with increased density stratification, the morphology of the zonal flows are somewhat insensitive to a change in density \citep{gastine2012dipolar,gastine2013zonal}.
The density anomaly that gives rise to the zonal gravitational moments is most dependent upon zonal winds, and so they should also not be strongly dependent on density stratification. This not the case for non-axisymmetric gravitational moments, however, so their properties could be changed as the circulation is modified by density stratification.
This should be the subject of future work.

\subsection{Magnetic fields}

Although numerous hypotheses exist for mechanisms that generate non-dipole-dominated magnetic fields and how the magnetic field morphology might be predicted via scaling laws (see \cite{soderlund2020underexplored} for a recent review), we focus here on the local Rossby number because $Ro_l>0.1$  predicts the occurrence of multipolar magnetic fields in Boussinesq dynamo models reasonably well \citep[e.g.,][]{christensen2010dynamo,gastine2012dipolar}.
As discussed in Section~\ref{sec:velocity_results}, the local Rossby numbers in our simulations have $Ro_l \gtrsim 0.3$; we, therefore, expect multipolar magnetic fields would be generated by these convective flows, not unlike those in the ice giants, which will be investigated in a follow-on paper. 

\subsection{Summary}

We have investigated whether the properties of Uranus (particularly in comparison to Neptune) are consistent with deep convection that is driven by compositional, rather than thermal, gradients.
While the relatively large internal heat flux of Neptune may be explained by primarily thermal convection, the low heat flux of Uranus may suggest a primarily chemical source of convection, which we have hypothesized results from the properties of mixtures of carbon-containing and oxygen- and nitrogen-containing compounds that comprise the planets and lead to the release of buoyant light elements in the deep interior.
In chemically driven cases, thermal contrasts across the interior are replaced, either partially or totally, by compositional contrasts, which may produce detectable effects such as changes to relative atmospheric abundances.
In most cases, zonal flows akin to those observed at Uranus and Neptune are generated, although the flow speeds are considerably stronger. 
In addition, buoyancy drives large convective cells that contribute to the transport of either heat or light elements to the surface, featuring larger heat fluxes in near the equator and in the polar regions, but suggesting that mass flux may become less variable with latitude than heat flux.
Furthermore, convection in the interior provides a means for sustaining a dynamo, and the dynamic similarities between thermal and compositional suggest that these magnetic fields should be non-dipolar, which may explain the similarities of the magnetic fields of Uranus and Neptune, despite thermal differences.

Current observations of atmospheric composition (particularly its vertical structure) are limited to the top few bars of the atmosphere, but future \textit{in situ} measurements, including those made by an atmospheric probe and gravity science, could improve our understanding of deep structure and dynamical processes, including the ice-to-rock ratio and mass fluxes predicted by our compositional convection models.
Furthermore, the decrease of diffusivity of mass, relative to heat, results in more locally concentrated parcels of fluid enriched in light elements that may result in detectable gravitational features that reflect the interior dynamics at spherical harmonic degrees above $n \geq 6$, which may be revealed by improved gravity science at the ice giants.

\section{Acknowledgments}

This work used the Extreme Science and Engineering Discovery Environment (XSEDE) Stampede2 at the University of Texas at Austin through allocation TG-PHY200059.  SLWM acknowledges support from NSF grant AST-1814772. KMS acknowledges support from NASA grant NNX15AL56G.

\bibliography{bibliography}{}
\bibliographystyle{aasjournal}

\end{document}